%% file: cspdata.tex
\documentclass[usenatbib,longnamesfirst,usedcolumn]{mn2e} 
\usepackage{epsfig}
\usepackage{lscape}  
\usepackage{longtable} 
\title[Gas fraction and the \MT\ relation]{The Birmingham-CfA 
cluster scaling project - I: gas fraction and the \MT\ relation}
\author[A. J. R. Sanderson et al.]
       {A. J. R. Sanderson$^{1,5}$ \thanks{E-mail: ajrs@astro.uiuc.edu}, 
          T. J. Ponman$^{1}$, A. Finoguenov$^{2,3}$, 
          E. J. Lloyd-Davies$^{1,4}$ \newauthor and M. Markevitch$^{3}$ \\
 $^{1}$School of Physics and Astronomy, University of
        Birmingham, Edgbaston, Birmingham B15 2TT, UK \\
 $^{2}$Max-Planck-Institut f\"ur extraterrestrische Physik,
        Giessenbachstra\ss e, 85748 Garching, Germany \\
 $^{3}$Harvard-Smithsonian Center for Astrophysics, 60 Garden Street, 
        Cambridge, MA 02138\\
 $^{4}$Department of Astronomy, University of Michigan, Ann Arbor, 
        MI 48109-1090, USA\\
 $^{5}$Department of Astronomy, University of Illinois, 
        1002 West Green Street, Urbana, IL 61801, USA
     \\}

 \date{Accepted 2002 ??.
      Received 2002 ??;
      in original form 2001 ??}

\pagerange{\pageref{firstpage}--\pageref{lastpage}}
\pubyear{2002}

\shortcites{arn01, blu84, bog89, boh02, cio91, dav01, ebe96, ebe98, eyl91, fai00,
  gou94, hel01, joh02, lew00, mar98b, mar99, mcn00, mua01, nev01,
  odrpack_ug, per98, pon94, pon96, sat00, sta92, tam01, vik99b}

\newcommand{\rmsub}[2]{\ensuremath{#1_{\mathrm{#2}}}} 
\newcommand{\srel}[2]{\mbox{\ensuremath{#1 - #2}}} 
\newcommand{\ASCA}{\emph{ASCA}}
\newcommand{\BSAX}{\emph{BeppoSAX}}
\newcommand{\cf}{{\textrm c.f.}}
\newcommand{\Chandra}{\emph{Chandra}}
\newcommand{\chisq}{\ensuremath{\chi^2}}
\newcommand{\cm}{\ensuremath{\mbox{~cm}}}
\newcommand{\eg}{{\textrm e.g.}}
\newcommand{\Einstein}{\emph{Einstein}}
\newcommand{\erg}{\ensuremath{\mbox{~erg}}}
\newcommand{\ergps}{\ensuremath{\erg \ps}}
\newcommand{\etal}{\mbox{{\textrm et al.\thinspace}}} 
\newcommand{\ewkT}{\rmsub{T}{ew}}
\newcommand{\fgas}{\rmsub{f}{gas}}
\def\h70{\rmsub{h}{70}} 
\newcommand{\ie}{{\textrm i.e.}}
\newcommand{\keV}{\ensuremath{\mbox{~keV}}}
\newcommand{\km}{\ensuremath{\mbox{~km}}}
\newcommand{\kmpspMpc}{\ensuremath{\km \ps \pMpc\,}}
\newcommand{\kpc}{\ensuremath{\mbox{~kpc}}}
\newcommand{\LB}{\rmsub{L}{B}}
\newcommand{\LT}{\srel{L}{\TX}}
\newcommand{\LX}{\rmsub{L}{X}}
\newcommand{\LXLB}{\mbox{\LX/\LB}}
\newcommand{\MEKAL}{\textsc{mekal}}
\newcommand{\Mpc}{\ensuremath{\mbox{~Mpc}}}
\newcommand{\Msol}{\rmsub{M}{\odot}}
\newcommand{\MT}{\srel{M}{\TX}}
\newcommand{\pcmsq}{\ensuremath{\cm^{-2}}}
\newcommand{\pMpc}{\ensuremath{\Mpc^{-1}}}
\newcommand{\ps}{\ensuremath{\s^{-1}}}
\def\R200{\rmsub{R}{200}} 
\newcommand{\rc}{\rmsub{r}{c}}
\newcommand{\ROSAT}{\emph{ROSAT}}
\newcommand{\RV}{\rmsub{R}{v}}
\newcommand{\s}{\ensuremath{\mbox{~s}}}
\newcommand{\Tbar}{\ensuremath{\overline{T}}}
\newcommand{\tcool}{\rmsub{t}{cool}}
\newcommand{\TX}{\rmsub{T}{X}}
\newcommand{\XMM}{\emph{XMM-Newton}}
\newcommand{\zf}{\rmsub{z}{f}}
\newcommand{\zobs}{\rmsub{z}{obs}}

\begin{document}

\maketitle

\label{firstpage}

\begin{abstract}

 \noindent We have assembled a large sample of virialized systems,
 comprising 66 galaxy clusters, groups and elliptical galaxies with high
 quality X-ray data. To each system we have fitted analytical profiles
 describing the gas density and temperature variation with radius, corrected
 for the effects of central gas cooling. We present an analysis of the
 scaling properties of these systems and focus in this paper on the gas
 distribution and \MT\ relation. In addition to clusters and groups, our
 sample includes two early-type galaxies, carefully selected to avoid
 contamination from group or cluster X-ray emission. We compare the
 properties of these objects with those of more massive systems and find
 evidence for a systematic difference between galaxy-sized haloes and groups
 of a similar temperature.
 
 We derive a mean logarithmic slope of the \MT\ relation within \R200\ of
 $1.84 \pm 0.06$, although there is some evidence of a gradual steepening
 in the \MT\ relation, with decreasing mass. We recover a similar slope
 using two additional methods of calculating the mean temperature.
 Repeating the analysis with the assumption of isothermality, we find the
 slope changes only slightly, to $1.89 \pm 0.04$, but the normalization is
 increased by 30 per cent. Correspondingly, the mean gas fraction within
 \R200\ changes from $(0.13 \pm 0.01)\h70^{-\frac{3}{2}}$ to $(0.11 \pm
 0.01)\h70^{-\frac{3}{2}}$, for the isothermal case, with the smaller
 fractional change reflecting different behaviour between hot and cool
 systems.  There is a strong correlation between the gas fraction within
 0.3\R200\ and temperature. This reflects the strong ($5.8\sigma$) trend
 between the gas density slope parameter, $\beta$, and temperature, which
 has been found in previous work.

 These findings are interpreted as evidence for self-similarity breaking 
 from galaxy feedback processes, AGN heating or possibly gas cooling. We 
 discuss the implications of our results in the context of a hierarchical 
 structure formation scenario.

\end{abstract}

\begin{keywords}
galaxies: clusters: general -- galaxies: haloes -- intergalactic 
medium -- X-rays: galaxies -- X-rays: galaxies: clusters
\end{keywords}

\section{Introduction}
\label{sec:intro}
The formation of structure in the Universe is sensitive to physical
processes which can influence the distribution of baryonic material, as
well as cosmological factors which ultimately govern the behaviour of the
underlying gravitational potential. By studying the properties of groups
and clusters of galaxies, it is possible to probe the physical processes
which shape the evolution and growth of virialized systems.

X-ray observations of the gaseous intergalactic medium (IGM) within a
virialized system provide an ideal probe of the structure of the halo,
since the gas smoothly traces the underlying gravitational potential.
However, this material is also sensitive to the influence of physical
processes arising from the interactions between and within haloes, which
are commonplace in a hierarchically evolving universe \citep[\eg][]{blu84}.
Even in relatively undisturbed systems, feedback from the galaxy members
can bias the gas distribution with respect to the dark matter in a way
which varies systematically with halo mass. N-body simulations
\citep[\eg][]{nav95} indicate that, in the absence of such feedback
mechanisms, the properties of the gas and dark matter in virialized haloes
should scale \emph{self-similarly}, except for a modest variation in dark
matter concentration with mass \citep{nav97}. Consequently, observations of
a departure from this simple expectation provide a key tool for
investigating the effects of \emph{non}-gravitational heating mechanisms,
arising from feedback processes.

There is now clear evidence that the properties of clusters and groups of
galaxies do not scale self-similarly: for example, the \LT\ relation in
clusters shows a logarithmic slope which is steeper than expected
\citep[\eg][]{edg91,arn99,fai00}. A further steepening of this slope is
observed in the group regime \citep[\eg][]{hel00}, consistent with a
flattening in the gas density profiles, which is evident in systems cooler
than 3--4\keV\ \citep{pon99}. Such behaviour is attributed to the effects
of non-gravitational heating, which exert a disproportionately large
influence on the smallest haloes. An obvious candidate for the source of
this heating is galaxy winds, since these are known to be responsible for
the enrichment of the IGM with heavy elements \citep[\eg][]{fin01b}.
However, active galactic nuclei (AGN) may also play a significant role,
particularly as there is some debate over the amount of energy available
from supernova-driven outflows \citep{wu00}. Recently, theoretical work has
also examined the role of gas cooling \citep[\cf][]{kni97}, which is also
able to reproduce the observed scaling properties of groups and clusters,
by eliminating the lowest entropy gas through star formation, thus allowing
hotter material to replace it \citep{mua01,voi01}.

Previous observational studies of the distribution of matter within
clusters have typically been limited by either a small sample size
\cite[\eg][]{dav95}, or have assumed an isothermal IGM \cite[\eg][]{whi95};
it appears that significant temperature gradients are present in many
\citep[\eg][]{mar98b}, although perhaps not all
\citep[\eg][]{whi00,deg02,irw00} clusters of galaxies. Another issue is the
restriction imposed by the arbitrary limits of the X-ray data; halo
properties must be evaluated at constant fractions of the virial radius
(\RV), rather than at fixed metric radii imposed by the data limits, in
order to make a fair comparison between varying mass scales. In this work,
we derive analytical expressions for the gas density and temperature
variation, which allow us to extrapolate these quantities beyond the limits
of the data. However, we are careful to consider the potential systematic
bias associated with this process. Our study combines the benefits of a
large sample with the advantages of a 3-dimensional, deprojection analysis,
in order to investigate the scaling properties of virialized haloes,
spanning a wide range of masses. In this work we have brought together data
from three large samples, comprising the majority of the suitable,
radially-resolved 3D temperature analyses of clusters. We include a large
number of cool groups in our analysis, as the departure from
self-similarity is most pronounced in haloes of this size: the
non-gravitationally heated IGM is only weakly captured in the shallower
potentials wells of these objects.

To further extend the mass range of our analysis, we include two
galaxy-sized haloes in our sample, in the form of an elliptical and an S0
galaxy. Galaxy-sized haloes are of great interest as they represent the
smallest mass scale for virialized systems and constitute the building
blocks in a hierarchically evolving universe. Great emphasis was placed on
identifying galaxies free of contamination from X-ray emission associated
with a group or cluster potential, in which they may reside, since this is
known to complicate analysis of their haloes \citep[\eg][]{mul98,hel00}.
The most well-studied galaxies are generally the first-ranked members in
groups or clusters, and it is known that such objects are atypical, as a
consequence of the dense gaseous environment surrounding them: the work of
\citet{hel01} has shown that brightest-group galaxies exhibit properties
which correlate with those of the group as a whole, possibly because many
of them lie at the focus of a group cooling flow. The study of
\citet{sat00} incorporated three ellipticals, but any X-ray emission
associated with these objects is clearly contaminated by emission from the
group or cluster halo in which they are embedded.

Throughout this paper we adopt the following cosmological parameters;
$H_{0}=70$\kmpspMpc and $q_{0}=0$. Unless otherwise stated, all quoted
errors are $1 \sigma$ on one parameter.

\section{The Sample}
\label{sec:sample}
In order to investigate the scaling properties of virialized systems, we
have chosen a sample which includes rich clusters, poorer clusters, groups
and also two early-type galaxies, comprising 66 objects in total. Sample
selection was based on two criteria: firstly, that a 3-dimensional gas
temperature profile was available. In conjunction with the corresponding
gas density profile, this allows the gravitating mass distribution to be
inferred. Secondly, we reject those systems with obvious evidence of
substructure, where the assumption of hydrostatic equilibrium is not
reasonable; it is known that the properties of such systems differ
systematically from those of relaxed clusters \citep[\eg][]{rit02}. This
also favours the assumption of a spherically symmetric gas distribution,
which is implicit in our deprojection analysis.

\onecolumn  
\input{target_table} 
\twocolumn 

By combining three samples from the work of Markevitch, Finoguenov and
Lloyd-Davies (described in detail in sections~\ref{ssec:MMdata},
\ref{ssec:AFdata} \&~\ref{ssec:clfitdata}, respectively) together with new
analysis of an additional six targets (also described in
section~\ref{ssec:clfitdata}), we have assembled a large number of
virialized objects with high-quality X-ray data. From these data, we have
derived deprojected gas density and temperature profiles for each object,
thus freeing our analysis from the simplistic assumption of isothermality
which is often used in studies of this nature. The large size of our sample
ensures a good coverage of the wide range of emission-weighted gas
temperatures, spanning 0.5 to 17\keV. Thus, we incorporate the full range
of sizes for virialized systems, down to the scale of individual galaxy
haloes. The redshift range is $z=0.0036$--0.208 (0.035 median), with only
four targets exceeding a redshift of 0.1. Some basic properties of the
sample are summarised in Table~\ref{tab:sample}.

As a number of systems are common to two or more of the sub-samples, we are
able to directly compare data from different analyses, allowing us to
investigate any systematic differences between the techniques employed. We
present the results of these consistency checks in
section~\ref{sec:checks}. The diverse nature of our sample, with respect to
the different methods used to determine the gas temperature and density
profiles, insulates our study to an extent from the bias caused by relying
on a single approach. However, we are still able to treat the data in a
homogeneous fashion, given the self-consistent manner in which the cluster
models are parametrized (see section~\ref{ssec:clmod}).

\section{X-ray Data Analysis}
\label{sec:xrayanal}
The X-ray data used in this study were taken with the \ROSAT\ PSPC and
\ASCA\ GIS \& SIS instruments. Although now superseded by the \Chandra\ and
\XMM\ observatories, these telescopes have extensive, publicly available
data archives and are generally well-calibrated. In addition, the PSPC and
GIS detectors have a wide field of view, which is essential for tracing
X-ray emission out to large radii, particularly for nearby systems, whose
virial radii can exceed one degree on the sky. The use of three separate
detectors, on two different telescopes, enhances the robustness of our
analysis, by reducing potential bias associated with instrument-related
systematic effects.

Since this work brings together data from separate samples, there is
considerable variation in the form in which those data were originally
obtained. This necessitated a supplementary processing stage to convert the
data into a unified format, in order to treat them in a homogeneous
fashion. In the case of the Finoguenov sample, analytical profiles were
fitted to deprojected gas density and temperature points (see
section~\ref{ssec:AFdata} for details); for the Markevitch sample it was
necessary to calculate the gas density normalization for such an analytical
function, from the fitted data (section~\ref{ssec:MMdata}). However, our
chosen model parametrization -- described below -- was fitted directly to
the raw X-ray data for the remaining systems, including the Lloyd-Davies
sample (further details of the data analysis are given in
section~\ref{ssec:clfitdata}).

\subsection{Cluster models}
\label{ssec:clmod}
In order to evaluate the gas temperature and density in a virialized
system, as well as derived quantities such as gravitating mass, at
arbitrary radii, we require a 3-dimensional analytical description of these
data. A core index parametrization of the gas density, $\rho(r)$, is used,
such that
\begin{equation}
\rho(r)=\rho(0)\left[1+\left(\frac{r}{r_{c}}\right)^{2}\right]^
 {-\frac{3}{2}\beta},
\label{eqn:3Dbeta}
\end{equation}
where $r_{c}$ and $\beta$ are the density core radius and index parameter,
respectively. The motivation for the use of this parametrization is
essentially empirical, although simulations of cluster mergers are capable
of reproducing a core in the gas density, despite the cuspy nature of the
underlying dark matter distribution \citep[\eg][]{pea94}. However, in the
absence of merging, N-body simulations offer no clear explanation for the
presence of a significant core in the IGM profile, even when the effects of
galaxy feedback mechanisms are incorporated \citep{met97}.

The density profile is combined with an equivalent expression for the
temperature spatial variation, described by one of two models; a linear
ramp, which is independent of the density profile, of the form
\begin{equation}
T(r)=T(0)-{\alpha}r,
\label{eqn:T_linear}
\end{equation}
where $\alpha$ is the temperature gradient. Alternatively, the temperature
can be linked to the gas density, via a polytropic equation of state, which
leads to
\begin{equation}
T(r)=T(0)\left[1+\left(\frac{r}{r_{c}}\right)^{2}\right]^
 {-\frac{3}{2}\beta \left(\gamma - 1\right)},
\label{eqn:T_polytropic}
\end{equation}
where $\gamma$ is the polytropic index and \rc\ and $\beta$ are as defined
previously.

Together, $\rho(r)$ and $T(r)$ can be used to determine the cluster
gravitating mass profile as, in hydrostatic equilibrium, the following
condition is satisfied
\begin{equation}
M_{grav}\left(r\right)=-\frac{kT\left(r\right)r}{G\mu 
 \rmsub{m}{p}}\left[\frac{\mathrm{d}\ln{\rho}}{\mathrm{d}\ln{r}}+\frac{\mathrm{d}\ln{T}}{\mathrm{d}\ln{r}}\right],
\label{eqn:M(r)}
\end{equation}
\citep{sar88}, where $\mu$ is the mean molecular weight of the gas and
\rmsub{m}{p} is the proton mass. This assumes a spherically symmetric mass
distribution, which has been shown to be a reasonable approximation, even
for moderately elliptical systems \citep{fab84}.

Since the X-ray emissivity depends on the product of the electron and ion
number densities, we parametrize the gas density in terms of a central
electron number density (\ie\ at $r=0$), assuming a ratio of electrons to
ions of 1.17. We base our inferred electron densities on the X-ray flux
normalized to the \ROSAT\ PSPC instrument, as there is a known effective
area offset between this detector and the \ASCA\ SIS and GIS instruments.
In those systems where the original density normalization was defined
differently, a conversion was necessary and this is described below.

Once the gravitating mass profile is known (from equation~\ref{eqn:M(r)}),
the corresponding density profile can be found trivially, given the
spherical symmetry of the cluster models. This can then be converted to an
overdensity profile, $\delta(r)$, given by
\begin{equation}
  \delta(r) \: = \: \frac{\rho_{tot}(r)}{\rho_{crit}},
\label{eqn:overdensity}
\end{equation}
where $\rho_{tot}(r)$ is the mean total density within a radius, $r$, and
$\rho_{crit}$ is the critical density of the Universe, given by 
$3H_{0}^{2} / 8\pi G$.

It is the overdensity profile which determines the virial radius (\RV) of
the cluster; simulations indicate that a reasonable approximation to \RV\ 
is given by the value of $r$ when $\delta(r)=200$ \citep[\eg][]{nav95} --
albeit for $\rho_{tot}(r)$ calculated at the redshift of formation, \zf,
rather than the redshift of observation, \zobs -- and we adopt this
definition in this work. Strictly speaking, the approximation $\RV=\R200$
is cosmology-dependent but, in any case, the implicit assumption $\zf =
\zobs$ is a greater source of uncertainty. In particular, there is a
systematic trend for the discrepancy between these two quantities to vary
with system size, in accordance with a hierarchical structure formation
scenario, in which the smallest haloes form first. The consequences of this
effect are addressed in section~\ref{ssec:rvir}. Given the local nature of
our sample, the assumed cosmology has little effect on our results. For
example, comparing the values of luminosity distance obtained for $q_{0}=0$
and $q_{0}=0.5$: the difference is less than 5\% for our most distant
cluster ($z = 0.208$), dropping to less than 2\% for $z < 0.1$ (\ie\ for
94\% of our sample).

Length scales in the cluster models are defined in a cosmology-independent
form, with the core radius of the gas density expressed in arcminutes and
the temperature gradient in equation~\ref{eqn:T_linear} measured in keV per
arcminute. The contributions to the cluster X-ray flux, in the form of
discrete line emission from highly ionized atomic species in the IGM, are
handled differently between the different sub-samples. However, in all
cases the gas metallicity was measured directly in the analysis and hence
this emission has, in effect, been decoupled from the dominant
bremsstrahlung component, which we rely on to measure the gas density and
temperature.

The key advantage of quantifying gas density and temperature in an
analytical form, is the ability to extrapolate and interpolate these and
derived quantities, like gas fraction and overdensity, to arbitrary radius.
Consequently, the virial radius and emission-weighted temperature can be
evaluated in an entirely self-consistent fashion, and thus we are able to
determine the above quantities at fixed fractions of \R200, regardless of
the data limits.

Clearly, where this extrapolation is quite large (\eg\ at \R200) there is
potential for unphysical behaviour in the gas temperature, which is not
constrained to be isothermal. This is particularly true when steep
gradients are involved (\ie\ large values of $\alpha$ in
equation~\ref{eqn:T_linear} or values of $\gamma$ very different from unity
in equation~\ref{eqn:T_polytropic}). A linear temperature parametrization
is most susceptible to unphysical behaviour as it can extrapolate to
negative values within the virial radius. To avoid this problem, we have
identified those linear $T(r)$ models where the temperature within \R200\ 
becomes negative. In each case the alternative, polytropic temperature
description was used in preference, where this was not already the
best-fitting model.

\subsection{Cooling flow correction}
\label{ssec:cfcorr}
The effects of gas cooling are well known to influence the X-ray emission
from clusters of galaxies \citep{fab94b}. Cooling flows may be present in
as many as 70 per cent of clusters \citep{per98} particularly amongst
older, relaxed systems, where merger-induced mixing of gas is not a
significant effect. Consequently we expect cooling flows to be common in a
sample of this nature, as we discriminate against objects with strong X-ray
substructure, which is most often associated with merger events. It is
possible to infer misleading properties for the intergalactic gas, both
spatially and spectrally, if the contamination from cooling flows is not
properly accounted for. Specifically, gas density core radii -- and,
consequently, the $\beta$ index in equation~\ref{eqn:3Dbeta}
\citep[see][for example]{neu99} -- can be strongly biased, as can the
temperature profile, particularly as central cooling regions have the
highest X-ray flux.

In all of the sub-samples the effects of central cooling were accounted for
in the original analysis using a variety of methods, which are described in
the appropriate sections below. The final cluster models therefore
parametrize only the `corrected' gas density and temperature profiles;
thus, we have extrapolated the gas properties inward over any cooling
region, as if no cooling were taking place at all.

\subsection{Markevitch sample}
\label{ssec:MMdata}
The sub-sample of Markevitch (hereafter `M sample') was compiled from
several separate studies and comprises spatial and spectral X-ray data for
27 clusters of galaxies \citep{mar98b,mar98,mar99,mar97,mar96}. Of these
datasets, 22 are included in our final sample, the remaining systems being
covered by one of the other sub-samples (the factors affecting this choice
are described in section~\ref{sec:checks}).

To measure the spatial distribution of the gas, X-ray images of the
clusters were fitted with a modified version of equation~\ref{eqn:3Dbeta};
under the assumption of isothermality, equation~\ref{eqn:3Dbeta} leads to
an equivalent expression for the \emph{projected} X-ray surface brightness,
$S$, given by
\begin{equation}
S(r)=S(0)\left[1+\left(\frac{r_{p}}{r_{c}}\right)^{2}\right]^
 {-3\beta+\frac{1}{2}},
\label{eqn:2Dbeta}
\end{equation}
in terms of projected radius, $r_{p}$ as well as the density core radius,
$r_{c}$, and index, $\beta$. This is a modified King function or isothermal
$\beta$-model \citep{cav76}. For all but one of the clusters, data from the
\ROSAT\ PSPC were used for the surface brightness fitting, as this
instrument provides greatly superior spatial resolution compared to the
\ASCA\ telescope (for Abell~1650, no PSPC pointed data were available and
an \Einstein\ IPC image was used instead).

Although strictly only appropriate for a uniform gas temperature
distribution, this approach is valid since, for the majority of the
clusters in this sub-sample, the exponential cutoff in the emission lies
significantly beyond the \ROSAT\ bandpass ($\sim$0.2--2.4\keV).
Consequently, the X-ray emissivity in this energy range is rather
insensitive to the gas temperature, and therefore scales simply as the
square of the gas density. These images were also used directly as models
of the surface brightness distribution in order to determine the relative
normalizations between projected emission measures in the different regions
for which spectra were fitted using \ASCA\ data.

Gas density data for this sub-sample were provided in the form of a King
profile core radius and $\beta$ index, as derived from PSPC data, using
equation~\ref{eqn:2Dbeta}. However, the density normalization was only
available in the form of a central electron number density for a small
number of clusters: Abell~1650 \& Abell~399 \citep{jon99} and Abell~3558,
Abell~3266, Abell~2319 \& Abell~119 \citep{moh99}. In the original
Markevitch analyses, density normalization data for the remaining systems
were taken from \citet{vik99}, in the form of values of the radius
enclosing a known overdensity with respect to the average baryon density of
the Universe at the observed cluster redshift. It was therefore necessary,
for this work, to convert these values into central electron densities, to
provide the necessary normalization component in the cluster models.

Radii of overdensity of 2000, $R\arcmin$, were taken from \citet{vik99} and
were combined with the gas density core radii, $r_{c}$, and $\beta$ indices
to determine the density normalization, $\rho(0)$, given that
\begin{equation}
\rho(0)\int_{0}^{R\arcmin}\left[1+\left(\frac{r}{r_{c}}
 \right)^{2}\right]^{-\frac{3}{2}\beta}4\pi r^{2}dr=
 \frac{4\pi R\arcmin^{3}}{3}2000\rho(\zobs),
\label{eq:MM_rho(0)}
\end{equation}
where $\rho(\zobs)$ is the mean density of the Universe at the observed
redshift of the cluster. The integration was performed iteratively using a
generalisation of Simpson's rule to a quartic fit, until successive
approximations differed by less than one part in $10^{8}$.

The fitted gas density and temperature data for the M sample were corrected
for the effects of central gas cooling in the original analyses: the
cluster models based on these data parametrize only the uncontaminated
cluster X-ray emission. This was achieved by excising a central region of
the surface brightness data in the original analysis and, for the
temperature data, by fitting an additional spectral component in the
central regions (where required), to characterise the properties of the
cooling gas flux. Full details of these methods can be found in
\citet{vik99} and \citet{mar98b}.

Temperature data for all the clusters in this sub-sample were provided in
the form of a polytropic index and a normalization evaluated at $2 r_{c}$
(as defined in equation~\ref{eqn:T_polytropic}). This radius was chosen as
it lay within the fitted data region (\ie\ outside of any excised cooling
flow emission) in all cases. These fits results are based on the
\emph{projected} temperature profile, but have been corrected for the
effects of projection. To construct cluster models, it was necessary to
calculate $T(0)$ from these normalization values, by re-arranging
equation~\ref{eqn:T_polytropic} and substituting $r=2r_{c}$ to give
\begin{equation}
T(0)=T(2r_{c})\left[5+\left[\frac{3}{2}\beta \left(\gamma-1\right)
 \right]\right].
\label{eq:MM_T(0)}
\end{equation}
These central normalization values were combined with the corresponding
polytropic indices and density parameters to comprise a 3-dimensional
description of the gas temperature variation. Errors on all parameters were
determined directly from the confidence regions evaluated in the original
analyses.

\subsection{Finoguenov sample}
\label{ssec:AFdata}
The sub-sample of Finoguenov (hereafter `F sample') comprises X-ray data
compiled from several sources, incorporating a total of 36 poor clusters
and groups of galaxies \citep{fin99,fin00,fin00b,fin01b} which were subject
to similar analysis. Of the corresponding fitted results, 24 were used in
the final sample, with the remainder taken from one of the other
sub-samples (the factors affecting this choice are described in
section~\ref{sec:checks}). A combination of \ROSAT\ and \ASCA\ SIS
instrument data was used to determine the spatial and spectral properties
of the X-ray emission respectively.

Values for the King profile core radius and index parameter were taken from
surface density profile fits (using equation~\ref{eqn:2Dbeta}) to PSPC
images of the clusters, with the exception of HCG~51 and MKW~9, where no
such data were available and a \ROSAT\ HRI and \Einstein\ IPC observation
were used respectively. A central region of the surface brightness data was
excluded for all systems, to avoid the bias to \rc\ and $\beta$ caused by
emission associated with central gas cooling. The best-fitting parameters were
used to determine the 3-dimensional gas density and temperature
distribution, via an analysis of \ASCA\ SIS annular spectra, by fitting
volume and luminosity-weighted values in a series of spherical shells,
allowing for the effects of projection. In this stage of the analysis the
central cooling region was included and an additional spectral component
was fitted to the innermost bins, allowing this extra emission to be
modelled. A regularisation technique was used to stabilise the fit by
smoothing out large discontinuities between adjacent bins.  Further details
of this method can be found in \citet{fin99}.

To generate cluster models for these objects, it was necessary to infer a
central gas density normalization, as well as an analytical form for the
temperature profile. Density normalization was determined by a core index
function (see equation~\ref{eqn:3Dbeta}) fit to the data points, using the
$\beta$ index and core radius values from the PSPC surface brightness fits.
This was achieved by numerically integrating equation~\ref{eqn:3Dbeta} (as
described in section~\ref{ssec:MMdata}) between the radial bounds of the
spherical shells used to determine the fit points, weighted by $r^{2}$ to
allow for the volume of each integration element. The core radii and
$\beta$ index were fixed at their previously determined values and
$\rho(0)$ was left free to vary.  A best fit normalization was then found
by adjusting $\rho(0)$ so as to minimize the \chisq\ statistic. Confidence
regions for $\rho(0)$ were determined from those values which gave an
increase in \chisq\ of one.  Fitting was performed using the MIGRAD method
in the \textsc{minuit} minimization library from CERN \citep{minuit_ug} and
errors were found with MINOS, from the same package. For the core radius
and $\beta$ index parameters, a fixed error of four per cent was assumed,
based on an estimate of the uncertainties in the surface brightness fitting
\citep{fin01}.

Since the original density points were measured in units of proton number
density, it was necessary to convert them to electron number density for
consistency between the cluster models. It was also necessary to allow for
a known effective area offset between the \ASCA\ SIS and \ROSAT\ PSPC
instruments. This adjustment amounts to a factor of 1.2 multiplication to
convert from proton number densities inferred using the former, to
equivalent values measured with the latter.

An analytical form for the gas temperature profile was obtained from a
mass-weighted (\ie\ density multiplied by the integration element volume,
using equation~\ref{eqn:3Dbeta}) fit to the 3-dimensional data points,
excluding the cooling component. For 5 of the coolest groups (IC~4296,
NGC~3258, NGC~4325, NGC~5129 \& NGC~6329), the cold component was not
sufficiently separated from the bulk halo contribution and so those central
bins that were affected were excluded from the analytical fit. The best fit
temperature values were subsequently found for both a linear and polytropic
description, again based on the \chisq\ criterion.

The parametrization which gave the optimum (\ie\ lowest) \chisq\ fit to the
data points was used, except where this gave rise to unphysical behaviour
in the model; for three systems (Abell~1060, HCG~94 \& MKW~4) the linear
T(r) model led to a negative temperature within \R200, when extrapolated
beyond the data region; in these cases a polytropic description was used in
preference.

\subsection{Lloyd-Davies \& Sanderson samples}
\label{ssec:clfitdata}
The sub-sample of Lloyd-Davies (hereafter `L sample') comprises 19 of the
20 clusters and groups of galaxies analysed in the study of \citet{llo00}
(Abell~400 was omitted as it is thought to be a line-of-sight superposition
of two clusters). Of the corresponding fitted results, 14 were used in the
final sample, with the remainder taken from either the M or F samples (see
section~\ref{sec:checks}). \ROSAT\ PSPC data were analysed for all the
objects, with data from the wider passband \ASCA\ GIS instrument included
to permit the analysis of certain hotter clusters.

To extend the sample to include individual galaxies and also to improve the
coverage at low temperatures, an additional six objects were
analysed -- four groups and two early-type galaxies (this sub-sample is
hereafter referred to as the `S sample'). The galaxy groups were drawn
from the sample of \citet{hel00} and were chosen as being fairly relaxed
and having high-quality \ROSAT\ PSPC data available. Cooler systems, in
particular, were favoured, in order to increase the number of low mass
objects in the sample. The extra objects include two early type galaxies;
an elliptical, NGC~6482 and an S0, NGC~1553. 

Genuinely isolated early-type galaxies are rare objects, given the
propensity for mass clustering in the Universe. In addition, finding a
nearby example of such a system, which possesses an extended X-ray halo
that has been studied in sufficient detail to measure $T_{X}(r)$, severely
limits the number of potential candidates. Although NGC~1553 lies close to
an elliptical galaxy of similar size (NGC~1549) there is no evidence from
the PSPC data of any extended emission not associated with either of these
objects, which might otherwise point to the presence of a significant group
X-ray halo (see section~\ref{sssec:N1553}). NGC~6482, by contrast, is a
large elliptical ($\LB \sim 6 \times 10^{10} \rmsub{L}{B\odot}$) which
clearly dominates the local luminosity function and which is embedded in an
extensive X-ray halo ($\sim$100\kpc). Its properties indicate that this is
probably a `fossil' group (see section~\ref{sssec:N6482}) and as such, its
properties are expected to differ from those of an individual galaxy halo.

The data reduction and analysis for the S sample was performed in a similar
way to the study of \citet{llo00} and a detailed description can be found
there. The method used involves the use of a spectral `cube' of data -- a
series of identical images extracted in contiguous energy bands -- which
constitutes a projected view of the cluster emission. A three dimensional
model of the type described in section~\ref{ssec:clmod} can be fitted
directly to these data in a forward fitting approach \citep{eyl91}, in
order to `deproject' the emission. The gas density and temperature are
evaluated in a series of discrete, spherical shells and the X-ray emission
in each shell is calculated with a \MEKAL\ hot plasma code \citep{mew86}.
The emission is then redshifted and convolved with the detector spectral
response, before being projected into a cube and blurred with the
instrument point spread function (PSF). The result can be compared directly
with the observed data and the goodness-of-fit is quantified with a maximum
likelihood fit statistic \citep{cas79}. The model parameters are then
iteratively modified, so as to obtain a best fit to the data.

The contributions to the plasma emissivity from highly ionized species, in
the form of discrete line emission, is handled by parametrizing the
metallicity of the gas with a linear ramp (assuming fixed, Solar-like
element abundance ratios), normalized to the Solar value. However, the
poorer spectral resolution of the PSPC requires that the metallicity be
constrained to be uniform where only \ROSAT\ data were fitted (as for all
six extra systems in the S sample). For those clusters where \ASCA\ GIS
data were additionally analysed in the L sample (denoted by a `+' in the
right-most column of Table~\ref{tab:sample}), the gradient of the
metallicity ramp was left free to vary.

The use of maximum likelihood fitting avoids the need to bin up the data to
achieve a reasonable approximation to Gaussian statistics: a process which
would severely degrade spatial resolution in the outer regions of the
emission, where the data are most sparse. The only constraint on spatial
bin size relates to blurring the cluster model with the PSF; a process
which is computationally expensive and a strongly varying function of the
total number of pixels in the data cube. Although the Cash statistic
provides no absolute measure of goodness-of-fit, differences between values
obtained from the same data set are \chisq-distributed. This enables
confidence regions to be evaluated, for determining parameter errors
\citep[\cf][]{llo00}.

For the S sample, two different minimization algorithms were employed to
optimise the fit to the data. A modified Levenberg-Marquardt method
\citep{bev69} was generally used to locate the minimum in the parameter
space. Although very efficient, this method is only effective in the
vicinity of a minimum and is not guaranteed to locate the global minimum.
In several cases this approach was unable to optimise the cluster model
parameters reliably and a simulated annealing minimization algorithm was
used \citep{gof94}. However, the disadvantage of this technique is the
computational cost associated with the very large number of fit statistic
evaluations required: once the global minimum was identified, the
Levenberg-Marquardt method was used to determine parameter errors, in an
identical fashion to \citet{llo00}.

In order to determine errors on derived quantities, such as gravitating
mass and gas fraction, we adopt the rather conservative approach of
evaluating the quantity using the extreme values permitted within the
confidence ranges specified by the original fitted parameters. However,
although this method tends to slightly overestimate the errors, as can be
seen from the intrinsic scatter in our derived masses in
section~\ref{ssec:MT}, it is not liable to introduce a systematic bias into
any weighted fitting of these data.

For those systems in the L and S samples where a cooling flow component was
fitted, a power law parametrization was used to describe the gas
temperature and density variations within the cooling radius (also a fitted
parameter). To avoid unphysical behaviour at $R=0$, these power laws were
truncated at 10\kpc, well within the spatial resolution of the instrument
(for NGC~1395 a cut-off of 0.5\kpc\ was used to reflect the much smaller
size of its X-ray halo).

\subsubsection{NGC 1553}
\label{sssec:N1553}
The X-ray spectra of elliptical galaxies comprise an emission component
originating from a population of discrete sources within the body of the
galaxy, as well as a possible component associated with a diffuse halo of
gas trapped in the potential well. The contributions of these different
spectral components vary according to the ratio of the X-ray to optical
luminosity of the galaxy (\LXLB) \citep{kim92}. Since we are interested
only in the X-ray halo of the systems in this work, we favour those
galaxies with a high \LXLB, where the emission can be traced beyond the
optical extent of the stellar population.

A 14.5ks PSPC observation was analysed, in which the S0 galaxy NGC~1553
appears quite far off axis, although within the `ring' support structure.
Some 2000 counts were accumulated in the exposure and the emission is
detectable out to a radius of 4.8 arcmin (21\kpc). Although its \LXLB, of
$1.53\times10^{-3}$, does not mark it out as a particularly bright galaxy,
its X-ray halo is clearly visible and uncontaminated by group or cluster
emission. In fact, this ratio is typical of non-group-dominant galaxies
\citep[\cf][]{hel01}. However, for this reason we expect a reasonable
contribution to the X-ray flux from discrete sources; \citet{bla01} have
recently found that diffuse emission only accounted for $\sim$84 per cent
of the total X-ray luminosity in the range 0.3--1\keV, based on a 34ks
observation with the ACIS-S detector on board the \Chandra\ telescope.

The PSPC data show evidence of central excess emission, which is adequately
described by a power law spectrum, blurred by the instrument PSF, with a
photon index consistent with unity. This was modelled as a separate
component, so as to decouple its emission from that of the halo.
\citet{bla01} find evidence of a central, point-like source which they fit
with an intrinsically absorbed disk blackbody model. The spatial properties
of the X-ray halo are not addressed in their analysis, but in any case the
emission is only partly visible, due to the small detector area of the
ACIS-S3 CCD chip.

\subsubsection{NGC 6482}
\label{sssec:N6482}
The elliptical galaxy NGC~6482 is a relatively isolated object, which has
no companion galaxies more than two magnitudes fainter within $1\;
\rmsub{h}{50}^{-1} \Mpc$. However, its X-ray luminosity is in excess of
$10^{42} \rmsub{h}{50}^{-2} \ergps$, which is very large for a single
galaxy. These properties classify this object as a `fossil' group -- the
product of the merger of a number of smaller galaxies, bound in a common
potential well \citep{pon94,mul99,vik99b,jon00}. Correspondingly, this
system is more closely related to a group-sized halo -- albeit a very old
one \citep[\cf][]{jon00} -- than to that of an individual galaxy. The X-ray
over-luminous nature of this galaxy ($\LXLB = 0.048$) implies that the vast
majority of the emission originates from its large ($\ga$100\kpc) halo,
with a negligible contribution from discrete sources.

Approximately 1500 counts were accumulated in an 8.5ks pointing with the
PSPC. During the fitting process it was found that there was a significant
residual feature in the centre of the halo, which may indicate the presence
of an AGN. It was not possible to adequately model this feature with either
a point-like or extended component and it was necessary to excise a central
region (radius 1.2 arcmin) of the data to obtain a reasonable fit. As a
result, the core radius was rather poorly constrained and hence was frozen
at its best-fitting value of 0.2 arcmin for the error calculation stage. In
addition, the hydrogen column could not be constrained and had to be frozen
at the galactic value ($7.89 \times 10^{20}$\pcmsq), as determined from the
radio data of \citet{sta92}.

\section{Consistency between sub-samples}
\label{sec:checks}
As a consequence of converting the data from the different sub-samples into
a uniform, \emph{analytical} format, we are able to adopt a coherent
approach in our analysis. By extrapolating the gas density and temperature
profiles, it is possible to determine the virial radius and mean
temperature (see below) self-consistently, and thus independently of the
arbitrary data limits. Of course, this process of extrapolation can
potentially introduce other biases, and this is discussed in
section~\ref{ssec:extrap_bias} below. In some systems, emissivity profiles
are affected by significant central cooling and we emphasize that in our
analysis we have eliminated this contaminating component in all of our
targets, in order to maintain consistency between the different
sub-samples. In this section we present the results of an investigation
into the consistency of our sample and the agreement between the different
analysis involved.

\begin{figure*}
  \centering \vspace{-2cm}
  \psfig{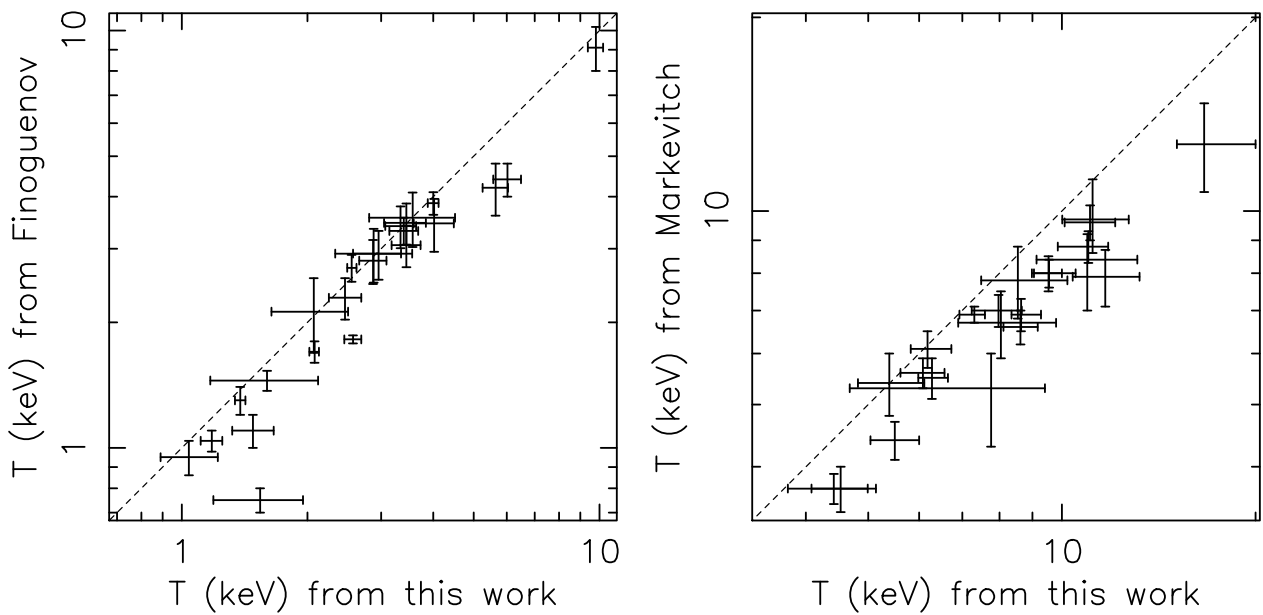} \vspace{-1.5cm}
\caption{\label{fig:T_comparison}
  Comparison of the emission-weighted temperatures from this work, with
  those from the original Finoguenov (left panel) and Markevitch (right
  panel) analyses. The dashed line indicates the locus of equality.}
\end{figure*}

Mean temperatures were calculated for each system, by averaging their gas
temperature profiles within 0.3\R200, weighted by emissivity and excluding
any cooling flow component (hereafter referred to as \ewkT; see column~5 in
table~\ref{tab:sample}). Fig.~\ref{fig:T_comparison} shows the
temperatures determined in this way, from the F \& M samples, compared to
the corresponding values taken from the original analyses. The F sample
(left panel) shows good agreement, although some discrepancy is expected,
due to differences in the prescription for obtaining \ewkT. However, two
clusters are clearly anomalous -- Abell~2670 and Abell~2597. The case of
A2670 is a known discrepancy, arising from an unusually high background in
the SIS observation. A2597 is an example of the complications of a large
cooling flow, which is more readily resolved in the SIS observation than
the GIS data. The values of \ewkT\ quoted in \citet{fin00b} for these
clusters are actually based on PSPC and GIS data respectively
(\citealt{hob97} and \citealt{mar98b}) and not on the SIS data analysed in
that paper. However, to maintain consistency we have used just the SIS data
to construct our model for these clusters.

The agreement between \ewkT\ values for the M sample (right panel) is less
good, but here differences are to be expected: the method used in this work
weights the temperature profile, between 0.3\R200\ and zero radius, by the
emissivity of the gas as determined by extrapolating $\rho(r)$ and $T(r)$
inwards from beyond the cooling flow region. In contrast, \citet{mar98b}
determine a flux weighting for their mean temperatures based on their
estimate of the emission measure from the \emph{non}-cooling gas within the
core region. For strong cooling flows, this gives a low weighting to the
central values of $T(r)$ compared to those values just outside the cooling
zone. Since almost all the systems in this sample have polytropic indices
in excess of one, their gas temperatures \emph{increase} towards the
centre, so the differences in the spatial weighting give rise to a
systematic difference between values of \ewkT\ determined with the two
methods. The overall effect of our analysis is actually to correct for the
consequences of gas cooling, rather than simply to exclude the contribution
from the cold component to the X-ray flux. This amounts to a simple
normalization offset -- the mean of the values of \ewkT\ from the M sample
is 18 per cent lower than that of the values determined in this work.

\begin{figure*}
  \psfig{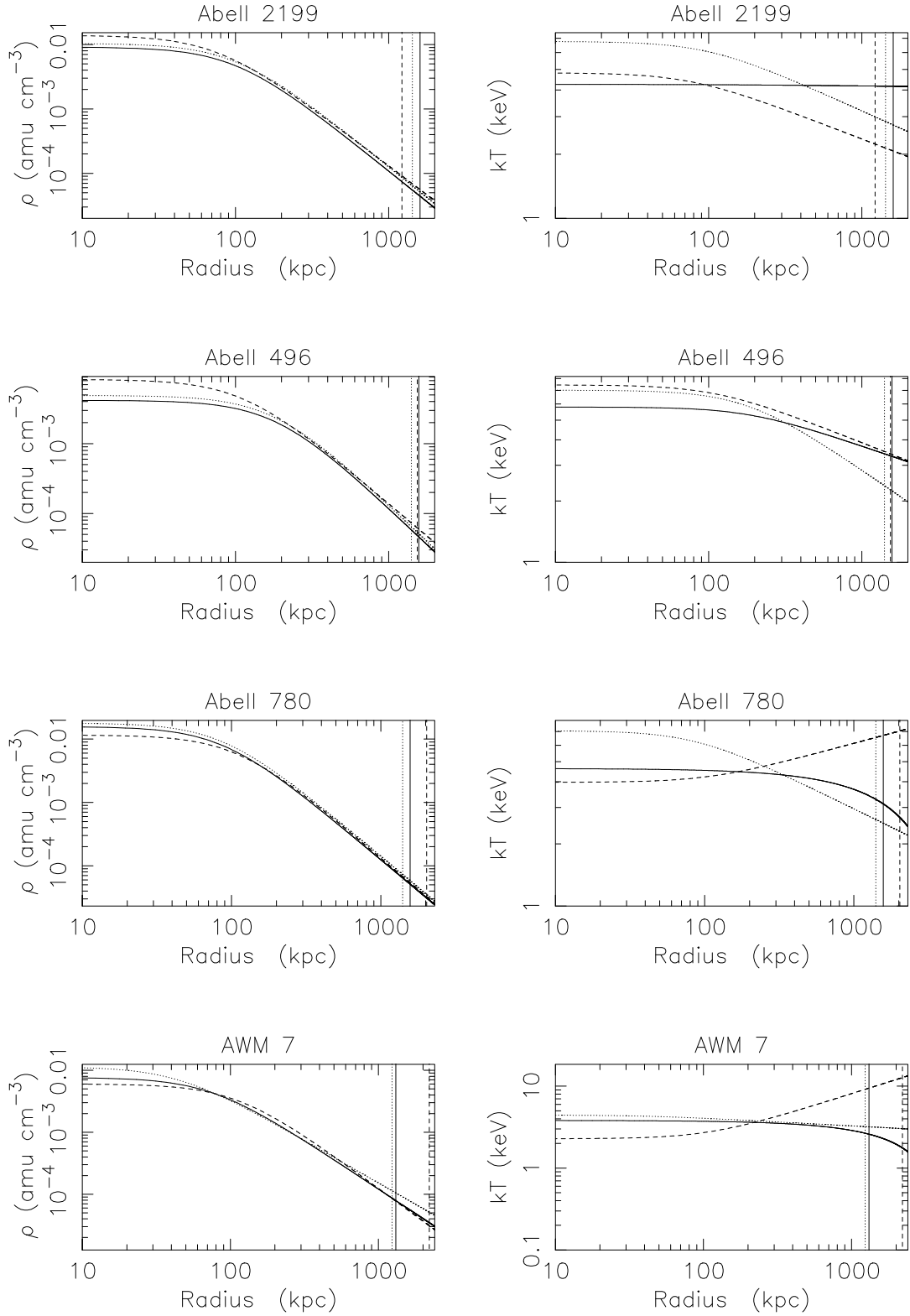}
\caption{\label{fig:4overlap_comparison}
  A comparison of the gas density and temperature profiles in four clusters
  common to the M (dotted lines), F (solid lines) and L (dashed lines)
  samples. The vertical lines mark the position of \R200\ for each of the
  different models.}
\end{figure*}

To assess the consistency between the different initial analyses in our
sample, we studied the models derived for four clusters which were common
to the M, F and L samples (Abell~2199, Abell~496, Abell~780 \& AWM~7),
providing a direct comparison of methods.
Fig.~\ref{fig:4overlap_comparison} shows the temperature and density
profiles for each of these systems -- in each plot the different lines
correspond to a different analysis result. It can be seen that the density
profiles show excellent agreement in all but the very central regions. At
the redshift of the most distant cluster ($z=0.057$, for A780), 1 arcmin
corresponds to roughly 60\kpc\ and hence these differences are confined to
the innermost parts of the data. Since these are all cooling flow clusters,
any discrepancies in the core can be attributed to differences in the way
the cooling emission is handled. In any case, the effects of these
discrepancies on the global cluster properties are small. The temperature
profiles show considerably more divergence, and for the clusters A780 and
AWM~7, the L sample temperature rises with radius, in contrast to the M and
F sample models. In the case of A780, data from a recent \Chandra\ analysis
\citep{dav01,mcn00} indicate that $T(r)$ does indeed show evidence of a
rise with radius within the inner $\sim$200\kpc\ in the ACIS-S detector
data, although the ACIS-I temperature profile exhibits a drop in the outer
bin, in the range 200--300\kpc.

The discrepancy between the temperature profiles of A780 and AWM~7 is
exacerbated by the rise with radius seen in the L sample models, which has
the compounding effects of increasing the size of \R200, as well as
steepening the gravitating mass profile. However, these clusters have two
of the most extreme rises in $T(r)$ of any system in our sample, and only 5
other systems show any significant increase in temperature with radius.
While it is clear that some clusters show evidence of a radially increasing
temperature profile in their central regions, it is unlikely that this will
continue out to the virial radius. This presents a fundamental problem for
a monotonic analytical profile, which must inevitably find a compromise: in
general the fit is driven by the central regions, which have a greater flux
weighting. In the case of A780, the difference in $T(r)$ leads to a factor
of 3 difference in the total mass within \R200, between the models,
although this discrepancy is reduced to 60 per cent for the mass within
0.3\R200. The corresponding effect on the gas fraction is also less severe,
since the total gas mass increases with \R200. However, for A496 -- whose
temperature profile is more typical of the systems in our sample -- the
agreement between the gravitating mass within \R200\ for the different
models is much better, varying by only 40 per cent.

\section{Final model selection}
In order to arrive at a single model for each system, we determined an
order of preference for the sub-samples, to choose between analyses, where
overlaps occurred. An initial selection was made on the basis of unphysical
behaviour in the models; the linear temperature parametrization is prone
to extrapolate to negative values within \R200, and so a number of models
were rejected on these grounds. Of the remaining overlaps, we
preferentially select those cluster models from the L sample, as this
represents the direct application of the model to the raw X-ray data and
hence should be the most reliable method. Application of this criterion
leaves just four remaining systems, where an overlap occurs between the F
and M samples. These were resolved on an individual basis; in each case the
analysis of the data which covered the largest angular area was chosen.
Since the ability to trace halo emission out to large radii is critical in
this study, this amounts to selecting the more reliable analysis. The
parameters for each of the final models are listed in
table~\ref{tab:sample}.

\section{Comparison with Chandra and XMM-Newton}
To provide a further cross-check on our results, we present here a
comparison of our temperature profiles with those measured using the
recently launched \Chandra\ and \XMM\ satellites. A2199 has been observed
with \Chandra\ and an analysis of these data has recently been presented by
\citet{joh02}. The projected temperature profile shows a increasing $T(r)$
from the core out to $\sim$2.2 arcmin (78\kpc), where it turns over and
flattens somewhat-- albeit with only 2 data points. This turnover radius is
identical to our own ``cooling radius'' as determined in the L sample
analysis (see column~11 of Table~\ref{tab:sample}). \Citeauthor{joh02}'s
deprojected $T(r)$ rises continually with radius, but is limited to the
central $\sim$4 arcmin of the cluster. 

An \XMM\ observation of A496 was recently analysed by \citet{tam01}. The
projected temperature profile rises from the core and turns over at roughly
3.5 (137\kpc) arcmin, in good agreement with our ``cooling radius'' of 3.44
arcmin. Although \citeauthor{tam01}'s deprojected $T(r)$ peaks at a
slightly larger radius (of $\sim$5 arcmin), it clearly indicates that the
temperature drops significantly beyond this point, in qualitative agreement
with our profile in Fig.~\ref{fig:4overlap_comparison}. However, closer
comparison with our results in the outer regions of the halo is hampered by
the fact that the data from both the A2199 and A496 observations are
restricted to the innermost $\sim$8\,arcmin.

In both these cases, the observed emission is dominated by flux from the 
central portion of the halo, where the temperature drops towards the core. 
Previously this phenomenon was thought to be a cooling flow, although 
recent higher quality data have revealed a lack of cool gas in this region 
\citep[see][and references therein]{boh02}. This component has been either 
modelled out or excluded from our analysis, to allow us to infer the
properties of the ambient IGM within this region, which accounts for the
discrepancy between the \Chandra\ and \XMM\ $T(r)$ and our profiles in
Fig.~\ref{fig:4overlap_comparison}. However, the potential for bias caused
by any cooling region is limited, since it is confined to a small central
part of the halo (the median ratio of the cooling radius to \R200\ in our
sample is 5 per cent). We note that the most distant cluster in our sample
(Abell~2163) has a sufficiently small angular size ($\R200\sim$11 arcmin)
to allow \Chandra\ to be able to observe most of its halo; \citet{mar01}
have shown that its temperature profile agrees reasonably well with the
\ASCA\ $T(r)$, which was the basis for the model we have used for this
cluster.

\section{Results}
\label{sec:results}

\subsection{Gas distribution}
\label{sec:res_gasdist}
\begin{figure*}
  \centerline{ \epsfig{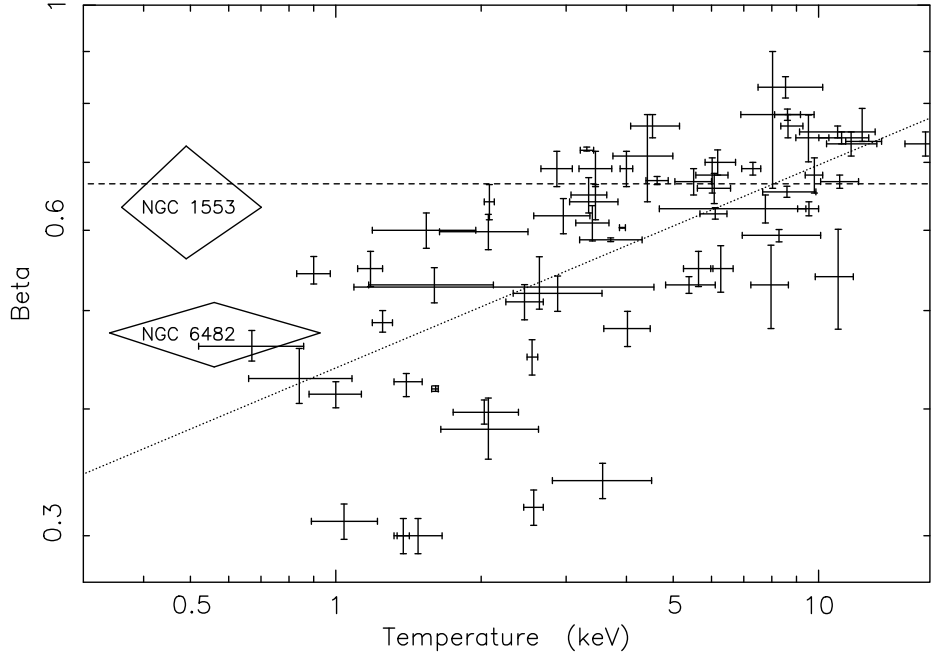} }
\caption{ \label{fig:beta-kT}
  The gas density slope parameter ($\beta$) as a function of system
  emission-weighted temperature. The diamonds represent the two galaxies in
  the sample. The dashed line indicates the canonical value of $\beta=2/3$
  and the dotted line is the best fit to the points, excluding the
  galaxies.}
\end{figure*}

Fig.~\ref{fig:beta-kT} shows the variation in the slope of the gas
density profile with emission-weighted temperature for the sample. It can
be seen that, for the hottest systems ($>$ 3--4\keV), $\beta$ is consistent
with the canonical value of 2/3 \citep[\eg][]{jon84}. However, below this
temperature the gas profiles become increasing flattened compared to
self-similar expectation, in agreement with the work of \citet{hel00}.
There is a strong correlation between $\beta$ and temperature as measured
by Kendall's K statistic, which gives a significance of 5.8$\sigma$.

\begin{figure*}
  \centerline{
    \epsfig{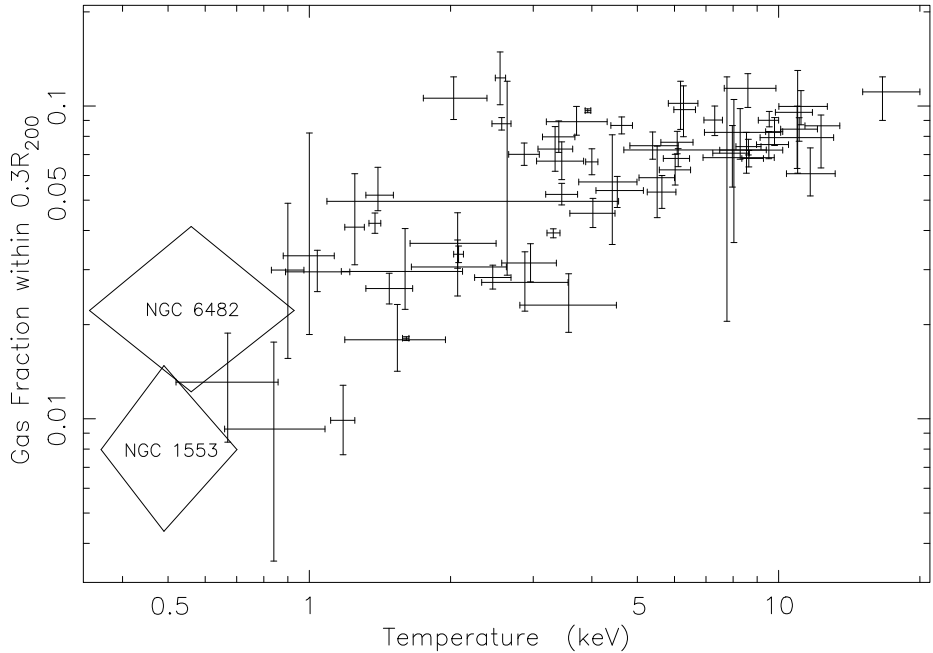} }
\caption{ \label{fig:fgas03RV-kT}
  Gas fraction within 0.3R$_{200}$ as a function of system temperature. The
  diamonds represent the two galaxies in the sample.}
\end{figure*}

Intriguingly, the galaxy (NGC~1553) and fossil group (NGC~6482) -- the
diamonds in Fig.~\ref{fig:beta-kT} -- seem to deviate from this general
trend. Although there are only two points, these are the coolest objects in
the sample and NGC~1553 in particular appears to have a value of $\beta$
more consistent with clusters than with groups of a similar temperature. We
will revisit this issue in the broader context of galaxy scaling properties
in section~\ref{ssec:galVsgrp}.

We fitted a straight line, in log space, to the points (both including and
excluding the two galaxies) using the \textsc{odrpack} software package
\citep{bog89,odrpack_ug}, to take account of parameter errors in both the X
and Y directions. The dotted line in Fig.~\ref{fig:beta-kT} shows the
best fit relation $\beta = (0.439\pm0.06) T^{0.20\pm0.03}$, excluding the
galaxies. The index is marginally consistent with the logarithmic slope of
$0.26\pm0.03$ found by \citet{hor99} for their literature-based sample,
spanning the range $\sim$1--10\keV. The flatter slope of our data reflects
the greater number of hotter clusters in our sample, where the relation
tends to flatten to approximately $\beta = 2/3$, indicated by the dashed
line in Fig.~\ref{fig:fgas03RV-kT}. The fit also matches the data points
from the simulations of \citet{met97}, which include the effects of galaxy
winds on the IGM, albeit with their points having a $\sim$25 per cent
higher normalization. A fit to the entire sample yields a flatter relation,
given by $\beta = (0.482\pm0.06) T^{0.15\pm0.03}$. Although the points seem
to be reasonably well described by a simple power law, there is a
considerable amount of intrinsic scatter in the data -- 80 per cent more
than would be expected from the statistical errors alone.

\begin{figure*}
  \centerline{ \epsfig{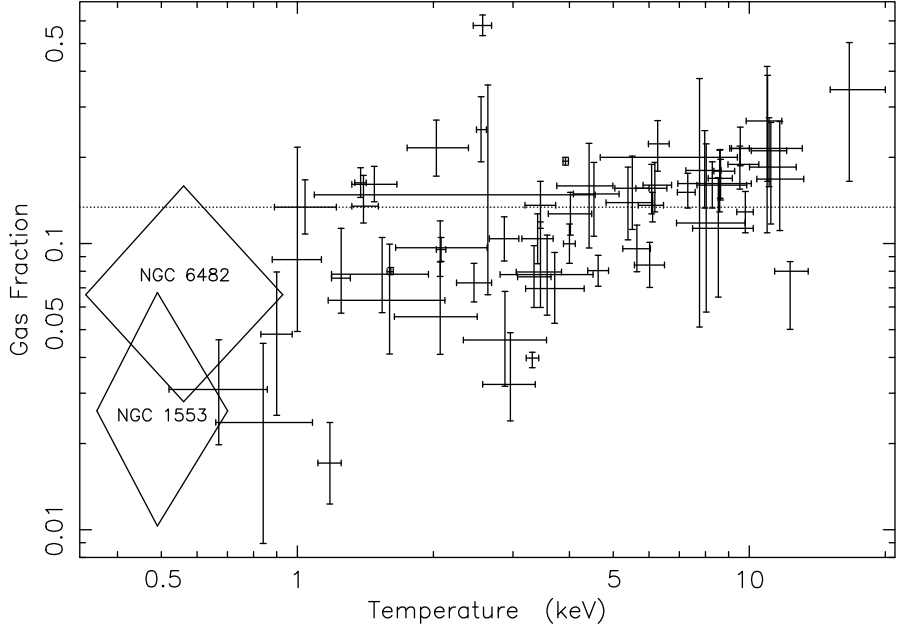} }
\caption{ \label{fig:fgasRV-kT}
  Gas fraction within R$_{200}$ as a function of system temperature. The
  diamonds represent the two galaxies in the sample and the dotted line
  shows the unweighted mean of the whole sample.}
\end{figure*}

The variation in $\beta$ is also reflected in the gas fraction (\fgas),
evaluated within a characteristic radius of 0.3\R200\ 
(Fig.~\ref{fig:fgas03RV-kT}). There is a clear trend (significant at the
6$\sigma$ level, excluding the two galaxies) for cooler systems to have a
smaller mass fraction of X-ray emitting gas. However, the galaxy NGC~1553
lies well below the general cluster relation, consistent with the coolest
groups, apparently at odds with its $\beta$ of approximately 2/3. This
behaviour is also evident in \fgas\ within \R200, shown in
Fig.~\ref{fig:fgasRV-kT}. By contrast, the fossil group NGC~6482 exhibits
gas fraction properties which are consistent with its $\beta$ -- \ie\
slightly above groups of a similar temperature in both cases. For the whole
sample, the behaviour of the gas fraction within \R200\ is only slightly
different from that within 0.3\R200; there still remains a strong
(5.4$\sigma$) trend, although there is some evidence of a levelling off
above $\sim$5\keV, above which the significance of a correlation drops to
3.2$\sigma$.

This can also be seen in the mean gas fraction within \R200\ for those
systems hotter than 4\keV, which gives $(0.163 \pm
0.01)\h70^{-\frac{3}{2}}$, as compared to $\fgas = (0.134 \pm
0.01)\h70^{-\frac{3}{2}}$ for the whole sample. Since the errors in the
evaluation of this quantity are dominated by systematic uncertainties, we
use an \emph{unweighted} mean \fgas, which is sensitive only to the
intrinsic scatter in the data. This behaviour suggests that virialized
objects may not be `closed systems', in that some of their gas might have
escaped beyond \R200, particularly for the coolest groups.  However, it
must be remembered that any effects of systematic extrapolation errors
could contribute to the observed trend.

\begin{figure*}
  \centerline{ \epsfig{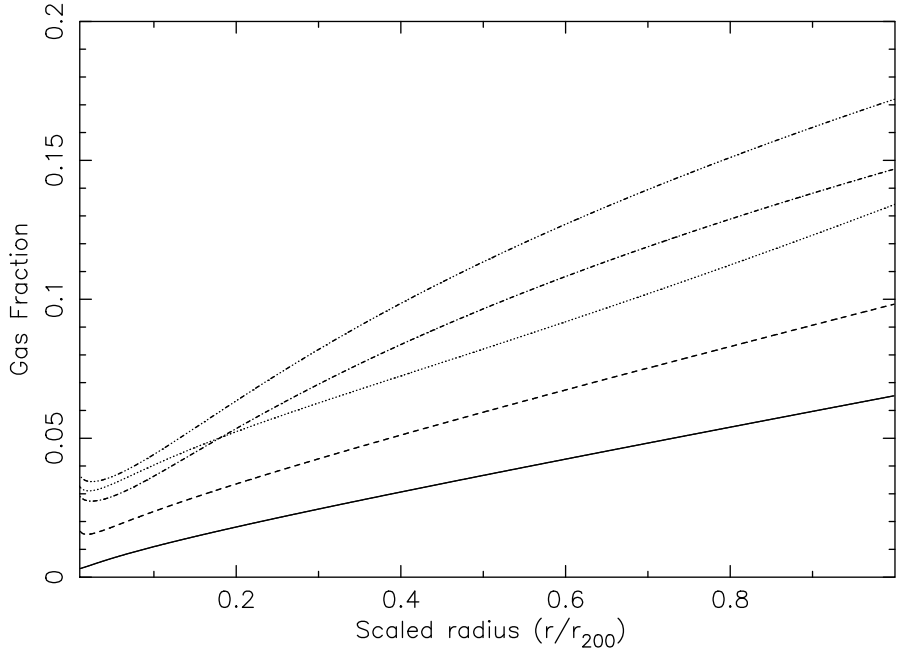} }
\caption{ \label{fig:fgas_profiles}
  Spatial variation of gas fraction within a given radius (normalized to
  R$_{200}$), grouped by system temperature. The solid line represents the
  coolest systems (including the two galaxies) (0.3--1.3\keV), increasing
  in temperature through dashed (1.3--2.9\keV), dotted (2.9--4.6\keV),
  dot-dashed (4.6--8\keV) and finally dot-dot-dot-dashed (8--17\keV).}
\end{figure*}

To understand the behaviour of the gas fraction across the sample,
Fig.~\ref{fig:fgas_profiles} shows how \fgas\ varies with radius, grouped
into five temperature bins for clarity. Beyond $\sim$0.2\R200, the profiles
lie in order of temperature such that, at a fixed radius, gas fraction
decreases as temperature decreases, mirroring the trend seen in
Fig.~\ref{fig:fgas03RV-kT}. This is essentially a simple normalization
offset and demonstrates that the effects of energy injection are more
pronounced in less massive (\ie\ cooler) systems, particularly below
$\sim$3--4\keV, as seen in Fig.~\ref{fig:fgasRV-kT}. The general trend is
for gas fraction to rise monotonically (beyond $\sim$0.03 \R200) with
radius from $\sim$0.02 in the core to around 18 per cent at \R200, for the
richest clusters ($kT > 8 \keV$). This behaviour demonstrates that the
distribution of the IGM is not similar to that of the dark matter, even in
the largest haloes, but is significantly more extended, as previously
reported \citep[\eg][]{dav95}.

\subsection{The \MT\ relation}
\label{ssec:MT}
Since the emission-weighted temperature reflects the depth of the
underlying potential well which retains the X-ray gas, a tight relation
between system mass and temperature is expected. It can be shown that, for
the case of simple self-similar scaling, $M \propto T^{3/2}$
\citep[see][for example]{moh97}. Observations generally reveal a steeper
relation, however, consistent with a breaking of self-similarity, as found
in other scaling relations (\eg\ \LT).

\ASCA\ temperature profiles, on which we rely in this work, have a
relatively large systematic uncertainty because of the wide mirror PSF. A
comparison of the \ASCA\ profiles from one of the subsamples used here
\citep[that of][]{mar98b} with recent \Chandra\ and \BSAX\ results appears
to confirm the temperature decline at large radii
\citep[\eg][]{dav01,mar01,nev01,deg02}. At the same time, \ASCA\ 
temperatures in the regions immediately adjacent to the central bins in the
cooling flow clusters appear to be systematically too high, although within
their uncertainties \citep[\eg][]{dav01,arn01,deg02}. Direct comparison is
limited to a few clusters at present.

It is important to correct for the effects of any central cooling flow when
calculating the characteristic temperature of a cluster, but is is not
obvious how best to achieve this. In our analysis below, we employ three
different methods extrapolating over, or excluding the central region, and
also weighting with the gas density rather than emissivity. The
justification for using these three different prescriptions for \Tbar\ is
as follows:
\begin{enumerate}
  \renewcommand{\theenumi}{(\roman{enumi})}
\item \textit{emission-weighted, extrapolating over CF}: this attempts to
  fully correct for the presence of a CF and provide an estimate of \Tbar\ 
  in the absence of cooling.
\item \textit{emission-weighted, excising cooling region} (radii of 
  excision are listed in column~11 of Table~\ref{tab:sample}): this method of
  calculating \Tbar\ more closely matches the CF-corrected, spectroscopic
  measurements which have been used frequently in previous work.
\item \textit{mass-weighted, extrapolating over CF}: this method gives
  values of \Tbar\ which are more naturally obtained from numerical
  simulations, and is less sensitive than emission weighting to the
  properties in the dense central core.
\end{enumerate}
We have applied these methods to derive \Tbar\ within two different radii:
\begin{enumerate}
  \renewcommand{\theenumi}{(\alph{enumi})}
\item 0.3\R200: the majority of our systems have X-ray emission
  detectable to at least this radius, which is typical of group detection
  radii.
\item \R200: this represents our nominal virial radius, and more
  closely matches the detection radii of rich clusters.
\end{enumerate}
We thus have six different methods of calculating \Tbar, including our
default method of emission-weighting $T(r)$ within 0.3\R200\ (\ie\ \ewkT,
described above and listed in Table~\ref{tab:sample}).

We have combined these temperature data with our gravitating mass
measurements (within both 0.3\R200\ and \R200, as appropriate) to give a
total of six \MT\ relations. Strictly speaking, the masses we derive should
be scaled by a factor of $(1+\zf)^{-3/2}$, to allow for the change in mean
density of the Universe with redshift. We have chosen to omit this
adjustment, since \zf\ is unknown, and the assumption $\zf = \zobs$ is
prone to systematically bias the results, as mentioned previously. We note,
however, that incorporating this correction actually makes very little
difference to the best-fitting parameters \citep{fin01}.

\begin{figure*}
  \vspace{-2.8cm} \psfig{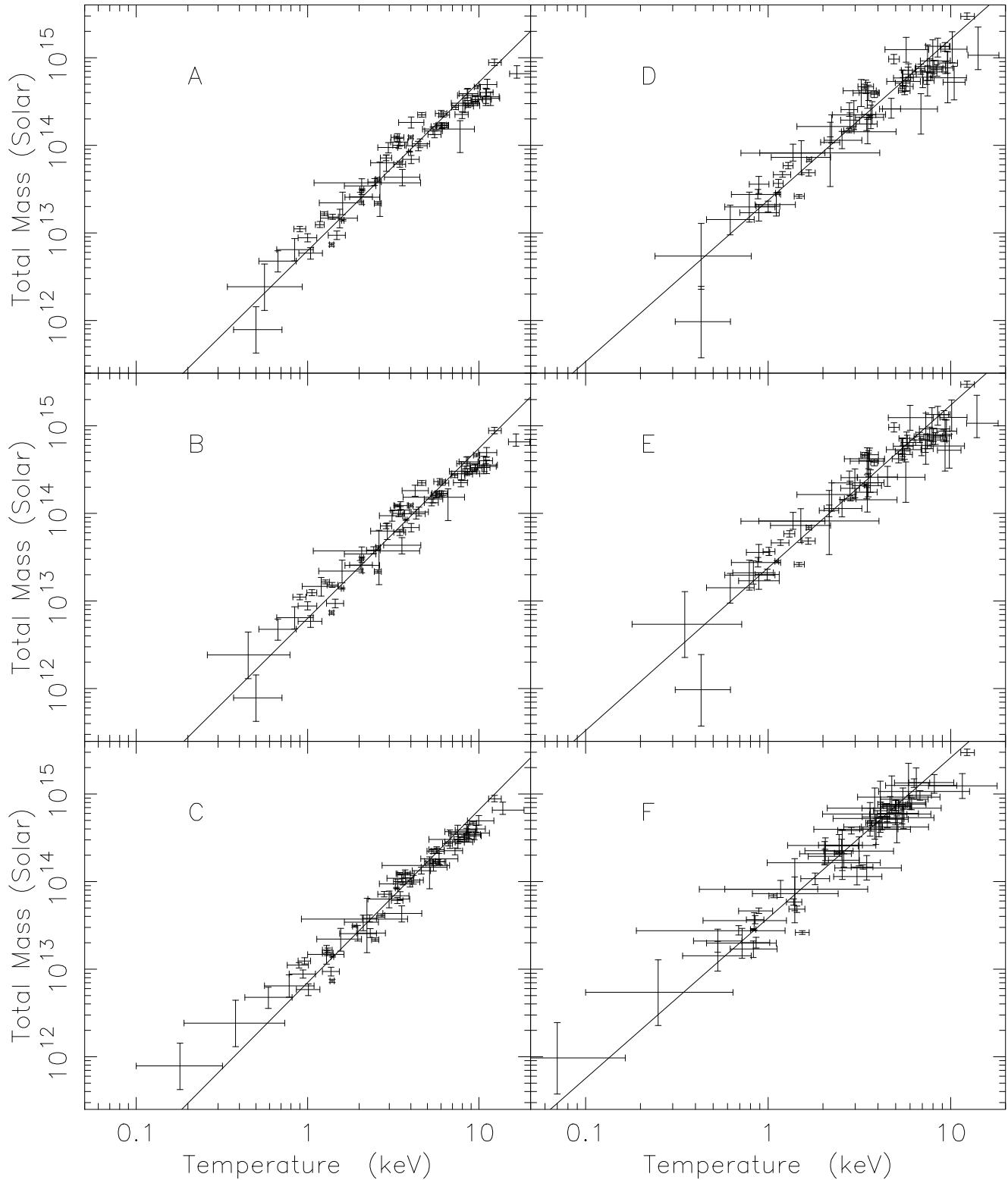}
  \vspace{-3.5cm}
\caption{\label{fig:6MT_mosaic}
  Total mass as a function of temperature for three different temperature
  prescriptions (rows) and measured within 0.3\R200\ (left column) and
  \R200\ (right column). In each case the solid line indicated the best fit
  power law. See upper half of Table~\ref{tab:MTfit} for further details.
  }
\end{figure*}

The set of \MT\ relations is plotted in Fig.~\ref{fig:6MT_mosaic},
together with the best-fitting power law in each case. The fitting was
performed in log space, using the \textsc{odrpack} software package, using
symmetrical errors in both axes derived from the half-widths of the
asymmetric errors on the original values. The upper section of
table~\ref{tab:MTfit} lists the parameters of the fit lines, together with
the corresponding scatter about the relation, normalized to that expected
from the statistical errors alone. A series of 1000 random realisations of
the data was generated by scattering each point away from the best-fitting
line, using the $1\sigma$ errors in both X and Y directions. The intrinsic
scatter was measured for the real data and for each simulated dataset, by
summing in quadrature the orthogonal distance of each point from the
best-fitting line. The real scatter was then normalized to the mean scatter
from all the realisations, to give the numbers quoted in column~5 of
table~\ref{tab:MTfit}. In each case, the level of scatter is fully
consistent with the errors, thus justifying the use of a weighted,
orthogonal distance regression to determine the best fit.

For the emission-weighted temperature and mass within 0.3\R200 (method A in
table~\ref{tab:MTfit}), we find a best-fitting relation of $\log{(M/\Msol)} =
(12.80\pm0.03) + (1.92\pm 0.06)\times \log{T}$ for the whole sample.
Exclusion of the two galaxies has a negligible effect on this result.
Excision of the cooling region (method B) leaves the normalization of the
\MT\ relation unchanged and increases the index only marginally, to
$1.94\pm0.06$. The use of mass-weighting to evaluate \Tbar\ (method C)
yields a best-fit which is consistent with those of methods A and B. It is
therefore clear that, within 0.3\R200, the \MT\ relation shows a
significantly steeper logarithmic slope than the self-similar prediction of
3/2. The agreement between the different methods for obtaining \Tbar\ 
demonstrates the robustness of this result. The behaviour of the \MT\ 
relation within \R200\ is similar but with a somewhat less steep slope: all
three measurements of \Tbar\ (methods D,E \& F) are consistent in producing
a best-fitting power law index of $\sim$1.84. The two emission-weighted methods
(D \& E) have identical normalizations, but the effect of using a
mass-weighting is to increase this value by $\sim$60 per cent. Although
yielding a flatter slope compared to the \MT\ relation within 0.3\R200,
this is still significantly steeper than the self-similar prediction.
Since $T(r)$ generally drops with radius, and more of the emission arises
at large radius in cooler systems, we expect \Tbar\ for the latter to drop
more as we move from 0.3\R200\ to \R200, hence flattening \MT.

The study by \citet{sat00} of 83 clusters, groups and galaxies observed
with \ASCA\ found a logarithmic slope of $2.04 \pm 0.42$, using total mass
within \R200\ together with temperatures determined by spectral fitting.
This value is more consistent with the slope for our data within 0.3\R200
than \R200, which may indicate that averaging \Tbar\ within 0.3\R200\ 
provides a closer match with spectroscopically measured temperatures, since
real X-ray data are rarely detectable out to \R200.  \citet{nev00} measure
a logarithmic slope of $1.79 \pm 0.14$, calculating total mass within an
overdensity of 1000 (\rmsub{M}{1000}), also using spectroscopically-derived
temperatures.  Their normalization, of $\log{M} = 13.15$, is intermediate
between our values for 0.3\R200\ and \R200, as expected for an overdensity
of 1000. The slope of $1.78 \pm 0.09$ found by \citet{fin01} for a sample
of 39 clusters (using \rmsub{M}{500}) is consistent with our relation
within \R200, and their normalization of $13.28 \pm 0.05$, lies slightly
below our own values within this radius.

While many X-ray studies appear to suggest that the slope of the \MT\ 
relation is steeper than the self-similar prediction of 1.5, it has been
suggested that this may an artefact of the analysis. \citet{hor99} measure
a slope of $1.78 \pm 0.05$ for a sample of 38 clusters, using the $\beta$
model to estimate masses. In contrast, they find a slope of $1.48 \pm 0.12$
for a smaller sample of 11 clusters, for which they have spatially resolved
temperature profiles. They attribute the discrepancy to the simplistic
assumption of isothermality (see section~\ref{ssec:iso}) and confirm the
apparently self-similar slope of the \MT\ relation with another sample of
27 clusters with virial mass estimates. However, the virial mass estimator
is known to be susceptible to bias from interloper galaxies and the
presence of substructure. In addition, the X-ray data for their 11 cluster
sample are taken from the literature, and are therefore expected to be
correspondingly heterogeneous.

The differences between the temperatures obtained with the different
methods can be gauged by studying the right-most two columns of
table~\ref{tab:MTfit}.  These show the mean and standard deviation of the
ratios obtained by dividing the values of \Tbar\ found with each of methods
B to F with the corresponding ones determined using method A. It can be
seen that the effect of excising the cooling region, as opposed to
extrapolating over it, results in an average 2 per cent decrease in \Tbar,
consistent with the general trend for $T(r)$ to increase towards the
centre. An even larger drop in \Tbar -- of 8 per cent -- is observed when
comparing the mass-weighted values (method C) with the baseline set (A),
although the spread of ratios is increased ($\sigma = 0.14$). By averaging
over the whole of \R200\ (D to F), the mean temperature decreases compared
to method A, by 8 and 9 per cent for methods D and E (emission-weighted),
respectively. The similarity between these mean ratios reflects the
proportionately smaller influence of the cooling region-excision when
integrating over the entire cluster volume. The mass-weighted \Tbar\ within
\R200\ shows an even greater drop, of 22 per cent (albeit with
$\sigma=0.41$), compared to method A. This is due to gas mass dropping
off less sharply than luminosity, lending greater weight to the outer
regions, where the gas temperature is generally lower.

The level of scatter in our \MT\ data is consistent with, or smaller than
the scatter expected just from statistical errors (depending upon the way
in which the temperature is weighted) -- \ie\ values of $\sim$0.5--1.0 in
column~5 of Table~\ref{tab:MTfit} -- suggesting that our error bounds are
somewhat conservative, as previously described (see
section~\ref{ssec:clfitdata}).  This conservative approach helps to allow
for extra sources of error -- for example, simulations have shown that
deviations from hydrostatic equilibrium introduce a 15--30 per cent rms
uncertainty into hydrostatic mass estimates \citep{evr96}. In any case, it
can be seen that a power law is not an ideal description of the data in
several of the plots in Fig.~\ref{fig:6MT_mosaic}. This may reflect the
dominance of systematic effects when extrapolating out to large radii, or
could indicate that the data follow a different functional form. Careful
inspection of Fig.~\ref{fig:6MT_mosaic} reveals some evidence for a
convex shape in a few cases, suggesting that the logarithmic slope steepens
\emph{gradually} from the cluster to the group regime. A convex \MT\ 
relation was predicted by the simulations of \citet{met97}, but only where
the input energy provided by galaxy winds is assumed to be fully retained
as thermal energy in the IGM: their simulations do not show this behaviour
in practice, as the extra energy is predominantly expended in doing work
redistributing the gas within the potential.

\input{MT_fit_table} 

More recently, \citet{dos02} have developed a purely analytical model,
which predicts a convex \MT\ relation of the form: $M =
M_{0}T^{3}\left[1+\frac{T}{T_{0}}\right]^{-3/2}$, with $T_{0} = 2\keV$.
This leads to a curve with a self-similar slope of 2/3 at the high mass,
smoothly increasing to an asymptotic value of 3 for $kT \rightarrow 0$.
Their model includes the effects of non-gravitational heating on the
pre-virialized IGM, as well as shock heating, and is able to reproduce the
observed \LT\ relation with a similar, curved fit to the data points.

Comparisons with mass measurements using data from the latest X-ray
missions are rather limited at present. However, a recent \Chandra\ study
by \citet{all01} has found \MT\ and \LT\ relations in agreement with the
predictions of self-similarity, albeit from a small sample of only six rich
clusters. This study is based on analysis of both X-ray data and
gravitational lensing information and finds good agreement between mass
estimates derived from the different methods.  \citet{all01} find a \MT\ 
logarithmic slope of $1.51 \pm 0.27$ within a radius of overdensity of
2500, which is approximately equivalent to 0.3\R200. However, their result
is not directly comparable to our \MT\ relations A, B \& C above, since the
overdensity profiles in our sample are not self-similar, so
\rmsub{R}{2500}\ actually corresponds to a different fraction of \R200\ for
each system.

\begin{figure*}
  \centerline{\epsfig{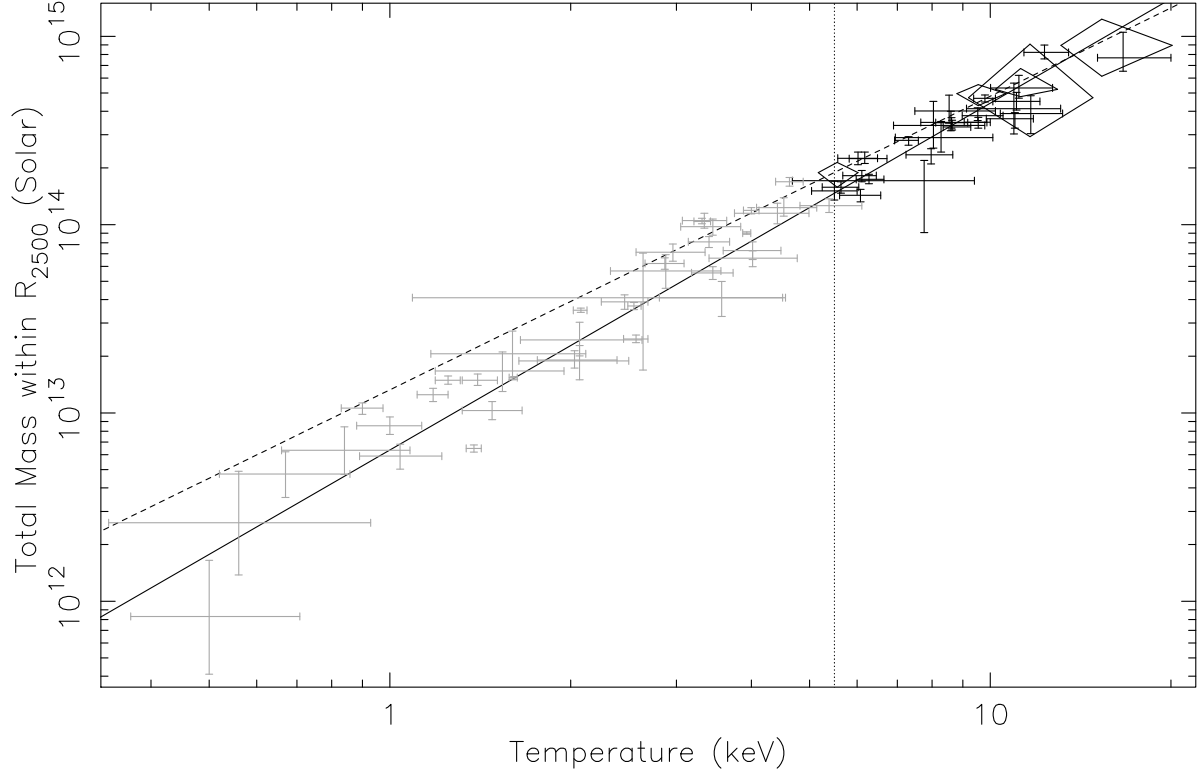} }
\caption{\label{fig:multi_MT_R2500}
  Total mass within \rmsub{R}{2500}\ as a function of emission-weighted
  temperature. The solid line is the best fitting power-law to the points
  above 5.5\keV\ (dotted vertical line), \ie\ excluding the grey points. The
  diamonds are the data of \citet{all01} and the dashed line is our
  best-fitting power-law to these data. See text for details.}
\end{figure*}

To permit a proper comparison, we have also derived masses within
\rmsub{R}{2500}\ and have fitted these data in an identical way to our
other \MT\ relations.  Fig.~\ref{fig:multi_MT_R2500} shows our results,
together with the five clusters from \citet{all01} which they use in their
\MT\ sample. It can be seen that their data points agree well with our
values at similar temperatures.  Fitting the five \citet{all01} clusters
using the same regression technique as we employed above, we find a
best-fitting slope of $1.56 \pm 0.16$ with an intercept of $\log{M} = 13.12
\pm 0.15$. If the sixth cluster from their sample (3C295) is included in
the fit, the best-fitting slope increases to $1.64 \pm 0.15$ and intercept
decreases to $13.04 \pm 0.14$.  This cluster was omitted by
\citeauthor{all01} as it was the only member of their sample without a
confirmed lensing mass estimate.  Fitting the clusters from our own sample
which are hotter than 5.5\keV\, for comparison with the \citet{all01}
analysis (their coolest cluster has $T=5.56$\keV), we find a logarithmic
slope of $1.84 \pm 0.14$, with a normalization of $\log{M} = 12.80 \pm
0.13$.  This is marginally consistent with the \citeauthor{all01} result.

If the difference in slope between the two samples of hot clusters is real,
it might be related to the dynamical state of the samples. The sample of
\citeauthor{all01} includes only the most relaxed clusters, where the
assumption of hydrostatic equilibrium has been independently verified by
lensing mass estimates. Our own sample is less well controlled, although we
have excluded objects which are clearly not in equilibrium.  On the other
hand, it is clear from Fig.~\ref{fig:multi_MT_R2500} that the shallower
slope from the \citeauthor{all01} data is a poor match to the relation for
cooler clusters, whilst the steeper slope of 1.84 fits rather well across
the entire temperature range.

Previous studies have suggested that the high and low mass parts of the
whole \MT\ relation may be characterised by power laws with different
slopes.  The cross-over temperature between the two regimes is typically
$\sim$3\keV\ \citep{fin01}. It is not obvious from our data that there is
such a break in the \MT\ relation, as has been found for the \LT\ relation
\citep[\eg][and references therein]{fai00}. \citet{fin01} find a steepening
of the logarithmic slope, from $1.48 \pm 0.11$ above 3\keV\ to $1.87 \pm
0.14$ below. However, this behaviour may simply be a manifestation of a
smooth transition with temperature, masked by a dearth of cool systems in
their sample, where the steepening slope is most apparent. More high
quality data of the type presented by \citet{all01}, but covering a wide
temperature range, will be required to establish whether the \MT\ is really
convex. What our results demonstrate clearly, is that either the relation
steepens towards lower mass systems, or its slope is substantially steeper
than 1.5.

\subsection{The effects of non-isothermality}
\label{ssec:iso}
To investigate directly the effects of neglecting spatial variations in gas
temperature, we have generated an additional set of \emph{isothermal}
models for our sample, \ie\ with $\alpha = 0$, for a linear $T(r)$, or
$\gamma = 1$, for a polytropic IGM. We have used the values of \Tbar\ 
already determined for the six different methods described above -- with
associated errors -- to define the constant value. These isothermal models
have then been subjected to an identical analysis to the original set, in
order to provide a fair comparison of results.

\begin{figure*}
  \centerline{\epsfig{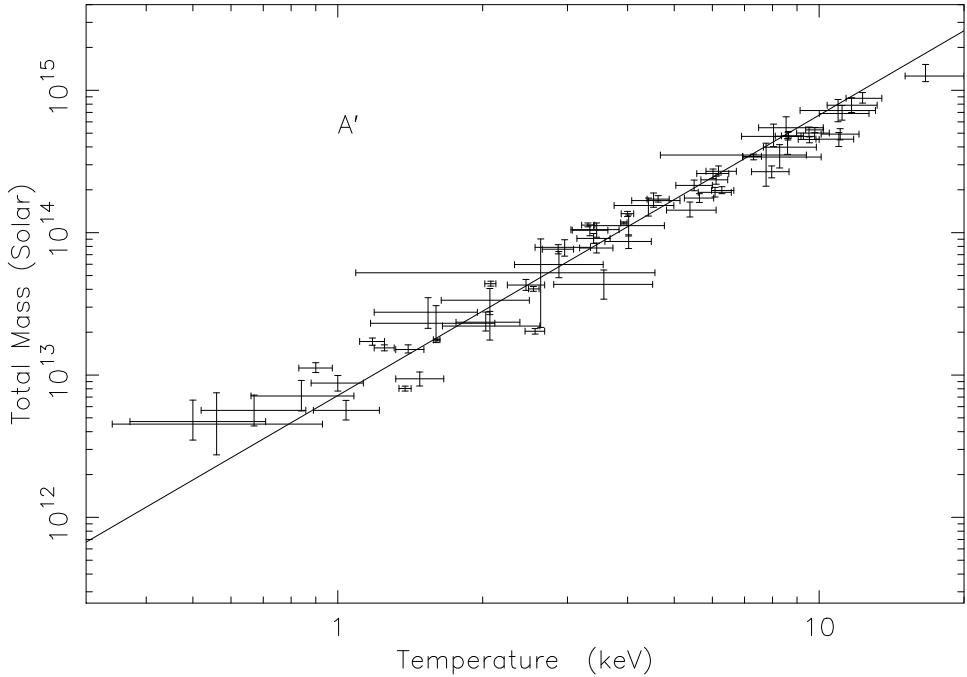} }
\caption{\label{fig:iso_MT1_03ewkT}
  Total mass as a function of emission-weighted temperature, evaluated
  within 0.3\R200, \textbf{for an isothermal IGM} (method A$^{\prime}$). The
  solid line represents the best-fitting power law. See lower half of
  Table~\ref{tab:MTfit} for further details.
        }
\end{figure*}

Fig.~\ref{fig:iso_MT1_03ewkT} shows the \MT\ relation for the isothermal
sample derived using temperatures from method A (referred to as
A$^{\prime}$).  It can be seen that the convex shape evident in panel A of
Fig.~\ref{fig:6MT_mosaic} is largely absent, and that a tighter relation
about the best-fitting line is observed. The parameters of this power law fit
are given in the lower half of Table~\ref{tab:MTfit}, together with
equivalent data for the other five isothermal \MT\ samples.  Within
0.3\R200\ the logarithmic slope increases marginally for the isothermal
models, but within the errors, for each of the three methods of measuring
\Tbar. However, for the two emission-weighted methods, the normalization
increases by $\sim$15 per cent, although it is unchanged for the
mass-weighted \Tbar.  Similar behaviour is observed for the \MT\ data
evaluated within \R200: the logarithmic slope is slightly steepened for
the isothermal case, and the normalization is increased -- for the
emission-weighted methods -- by $\sim$30 per cent. However, the
mass-weighted normalization \emph{decreases} by 15 per cent, compared to
the non-isothermal models.

It is clear from this that the assumption of isothermality leads to an
\emph{overestimate} of the total mass within \R200, when an
emission-weighted method is used to calculate \Tbar. A similar conclusion
was reached by \citet{hor99}, for a sample of 12 clusters, who found that
isothermality overestimated the mass by a factor of 1.7 -- a result
confirmed by \citet{neu99}. The latter authors found that the cumulative
mass within a given radius for an isothermal cluster is significantly
steeper than that of a cluster with a polytropic index of 1.25 (a value
typical of the systems in our sample -- see Table~\ref{tab:sample}), with
the intersection of the two occurring at $\sim$0.35\R200. Consequently, the
isothermal assumption over-predicts the mass for 96 per cent of the cluster
volume.

Neglecting temperature gradients in the IGM appears to have little or no
effect on the logarithmic slope of the \MT\ relation and, once again, the
observed slopes are in good agreement between the three different methods
of calculating \Tbar. This is in contrast to the prediction of
\citet{hor99}, who suggested that the assumption of isothermality leads to
a steepening in the \MT\ slope, which would otherwise be self-similar (\ie\
3/2). However, they base this conclusion on an analysis of a small sample
(12 systems), with data drawn from a number of different sources in the
literature. We also find that the rms scatter about the best fit \MT\ 
relations is significantly reduced in our isothermal models, and fully
consistent with that expected from the statistical errors. We conclude that
a power law seems to provide a good description of the \MT\ relation
\emph{for an isothermal IGM}.

\begin{figure*}
  \vspace{-3.5cm} \centerline{
    \epsfig{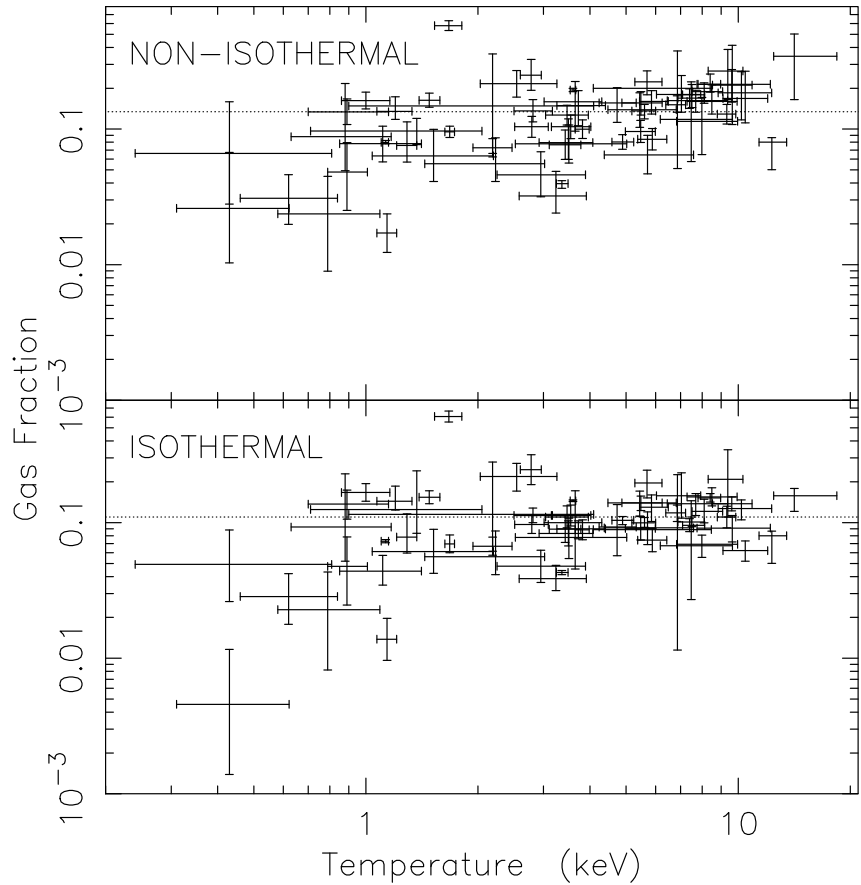} } \vspace{-2.5cm}
\caption{ \label{fig:2fgas_mosaic}
  \textit{Upper panel:} Gas fraction within \R200\ as a function of
  emission-weighted temperature (within \R200 -- method `D').
  \textit{Lower panel:} Gas fraction within \R200\ \textbf{for an
    isothermal IGM}. The dotted lines show the unweighted mean of the whole
  sample.  }
\end{figure*}

The overestimation of the total mass for the isothermal case leads to a
corresponding underestimation in the total gas fraction within \R200, shown
in Fig.~\ref{fig:2fgas_mosaic}. The unweighted mean gas fraction for the
whole sample is $(0.110 \pm 0.01)\h70^{-\frac{3}{2}}$, as compared to
$(0.134 \pm 0.01)\h70^{-\frac{3}{2}}$ for the non-isothermal case. It can
also be seen that the scatter about the mean is lower for the isothermal
case, although the apparent drop at $\sim$1\keV, seen in
Fig.~\ref{fig:2fgas_mosaic}, is still noticeable. The most obvious
outlier on this graph is the galaxy NGC~1553 (the left-most point). For
this system, the isothermal model results in a significantly lower \fgas,
which greatly increases its distance from the sample mean.

\subsection{Virial radius}
\label{ssec:rvir}
The precise location of the outer boundary of a virialized halo is
difficult to quantify and is very rarely directly observable. The virial
radius is dependent on the mean density of the Universe when the halo was
formed, as well as the adopted cosmology \citep{lac93}. Clearly it is
important to be able to define this quantity reliably, since we assume that
self-similar haloes will have identical properties when scaled by \RV. The
radius enclosing a mean overdensity of 200 (\R200) is proportional to \RV\ 
in any given cosmology -- and lies within \RV\ for all reasonable
cosmologies \citep{bry98} -- and scales in a identical way \citep{nav95}.
However, previous studies have not always been able to determine \R200, and
so have relied on other means to estimate this quantity. A tight
relationship between \ewkT\ and \RV\ (and hence \R200) is expected, as both
these quantities reflect the depth of the gravitational potential well in a
virialized halo; self-similarity predicts that $\RV\ \propto \sqrt{\ewkT}$
(\cf\ the size-temperature relation, \citealt{moh97}). This proportionality
has been confirmed in ensembles of simulated clusters, which provide a
value for the normalization in the relation. One such example is the work
of \citet{nav95}, who deduce that
\begin{equation}
\R200=0.813\left(\frac{T}{\rm{keV}}\right)^{\frac{1}{2}}
 (1+z)^{-\frac{3}{2}} \h70^{-1}\Mpc .
\label{eqn:NFW_kT_h70}
\end{equation}
However, their simulations only included adiabatic compression and shock
heating, and did not allow for the effects of energy injection.

\begin{figure*}
  \centerline{ \epsfig{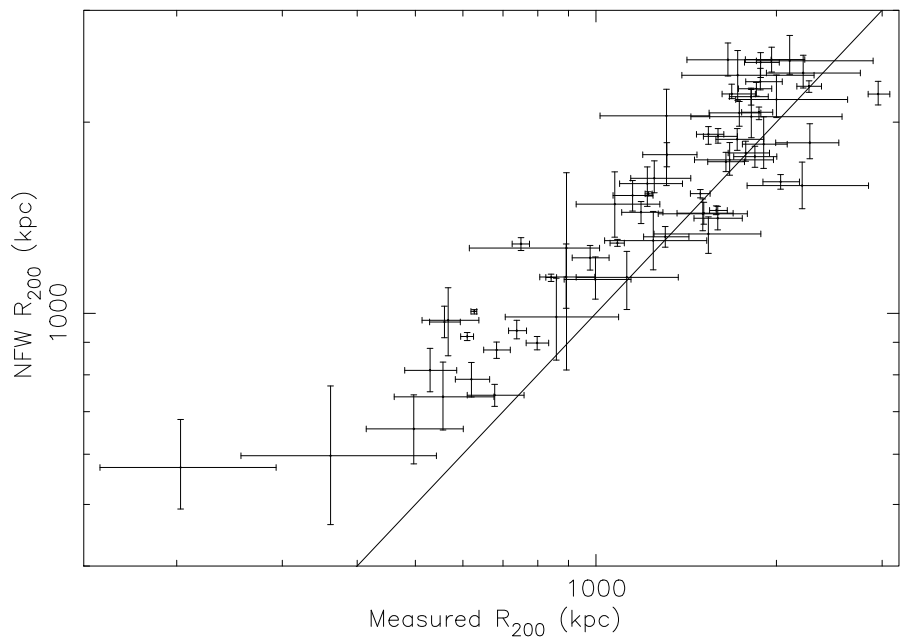} }
\caption{ \label{fig:R200vsNFW}
  Predicted \R200\ from the NFW formula (equation~\ref{eqn:NFW_kT_h70})
  plotted against measured \R200. The solid line indicates the line of
  equality.}
\end{figure*}

The correspondence between our values of \R200\ as determined from the
overdensity profile (listed in Table~\ref{tab:sample}) and those calculated
with equation~\ref{eqn:NFW_kT_h70} is shown in Fig.~\ref{fig:R200vsNFW}.
It can be seen that there is significant deviation from the locus of
equality between these quantities, marked by the solid line. The largest
discrepancy is observed in the smallest haloes, indicating that the NFW
equation significantly over-predicts \R200\ in these systems (the effect of
extrapolation bias is addressed in section~\ref{ssec:extrap_bias}). This is
to be expected, given that \ewkT\ for these objects is most likely to be
susceptible to bias from non-gravitational heating. To explore the reasons
for the disagreement between the two methods for calculating \R200, we have
examined the role of temperature gradients as the source of the scatter,
given their importance in calculating the gravitating mass (see
equation~\ref{eqn:M(r)}). We have defined a simple, quantitative measure of
the departure from isothermality, which, as has already been seen, can
exert a significant influence on scaling properties
(section~\ref{ssec:iso}). We use the ratio $T(0)/T(0.3\R200)$, as this is
very sensitive to the presence of a temperature gradient, and the two
distances involved bracket the region of interest used to calculate \ewkT.

\begin{figure*}
  \centerline{ \epsfig{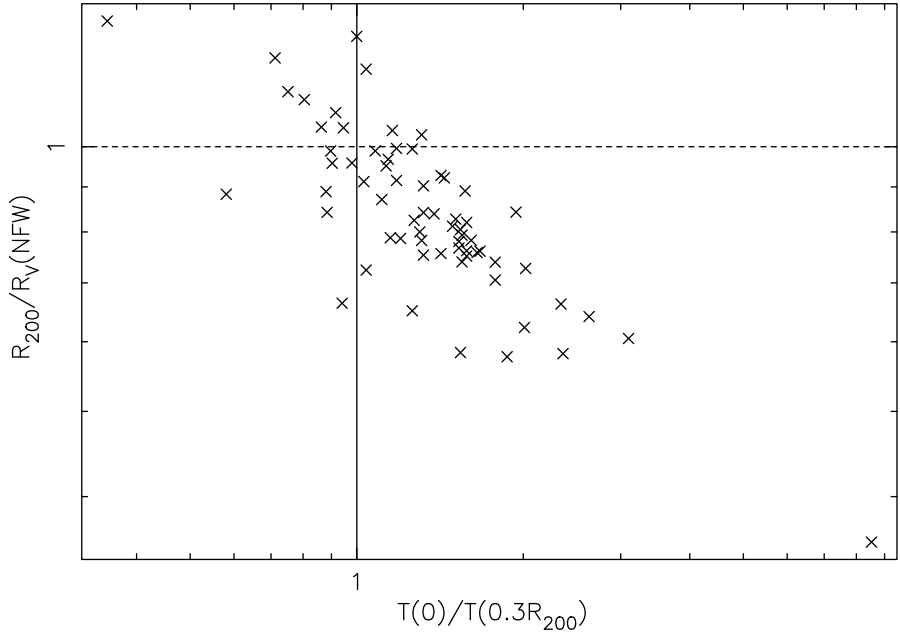} }
\caption{ \label{fig:residual_NFW}
  The ratio between measured \R200\ and \R200\ from the NFW formula, as a
  function of the ratio between $T(r)$  at $R = 0$ and $R= 0.3\R200$.
  Error bars have been omitted for clarity. The lines of equality on both
  axes are marked; the solid line represents the locus of isothermality.}
\end{figure*}

The relationship between this quantity and the ratio of the measured \R200\ 
divided by the NFW predicted value, is shown in
Fig.~\ref{fig:residual_NFW}. There is clearly a strong anti-correlation
between these quantities, significant at the $6.9\sigma$ level. Even with
the two most extreme points removed (the left and right-most points on the
graph), the significance of the relation drops only slightly, to
$6.4\sigma$.  It can be seen that the most isothermal systems (clustered
around the solid vertical line) scatter around the line of equality between
the two measurements of \R200. This demonstrates that the NFW formula is
valid only for nearly isothermal haloes, and that it otherwise over-predicts
\R200\ for the most common case of a radially decreasing temperature
profile.

\subsection{Extrapolation bias}
\label{ssec:extrap_bias}
Our analysis relies on the validity of extrapolating analytical profiles
fitted to an inner region of the data, in order to compensate for the
emission which is undetected. However, since the extrapolation is, in
general, greater for smaller systems, there is a potential for introducing
a systematic bias in our fitting. This is particularly true of the slope of
the gas density profile, which is best constrained by the emission from
outer regions of the X-ray halo, and which has been found to vary
significantly with temperature (see section~\ref{sec:res_gasdist}). The
work of \citet{vik99} has shown that there is evidence of a slight
steepening of the gas density logarithmic slope with radius. Such behaviour
would naturally lead to lower values of $\beta$ being inferred for cooler
systems, since their smaller haloes would be detectable to a smaller
fraction of \RV, compared to those of more massive clusters. This effect
could therefore explain part of the observed correlation between $\beta$
and \ewkT. To explore the effects of extrapolation on our data, we have
performed a series of fits to the surface brightness profiles of two
clusters (Abell~1795 \& Abell~2029), investigating the role of the outer
radius of the fitted data in constraining the fit parameters.

These clusters were selected as they are rich systems (and hence relatively
unaffected by energy injection), with high quality data ($\sim$100,000 and
30,000 counts, respectively), that cover a fairly large angular extent.
This allows us to trace the emission to a large fraction of \RV\ and means
we can analyse a spatial subset of the data, without approaching the
resolution limits of the instrument. In addition, we have chosen systems
which have cooling flow emission confined to as small a region as possible,
compared to the gas halo core size, so as to minimize the bias this
contamination can have on our results.

Since we are aiming in this section only to explore the behaviour of the
gas density logarithmic slope as a function of radius, we have adopted a
different approach to that described in section~\ref{ssec:clfitdata}.
Rather than applying a full deprojection analysis to the data, we have
fitted 1-dimensional, azimuthally-averaged, surface brightness profiles for
each cluster. This method permits a much more direct investigation of the
gas density index in the outer regions of the halo and allows us use a
quantitative measure of goodness-of-fit, based on the \chisq\ criterion.

\begin{figure*}
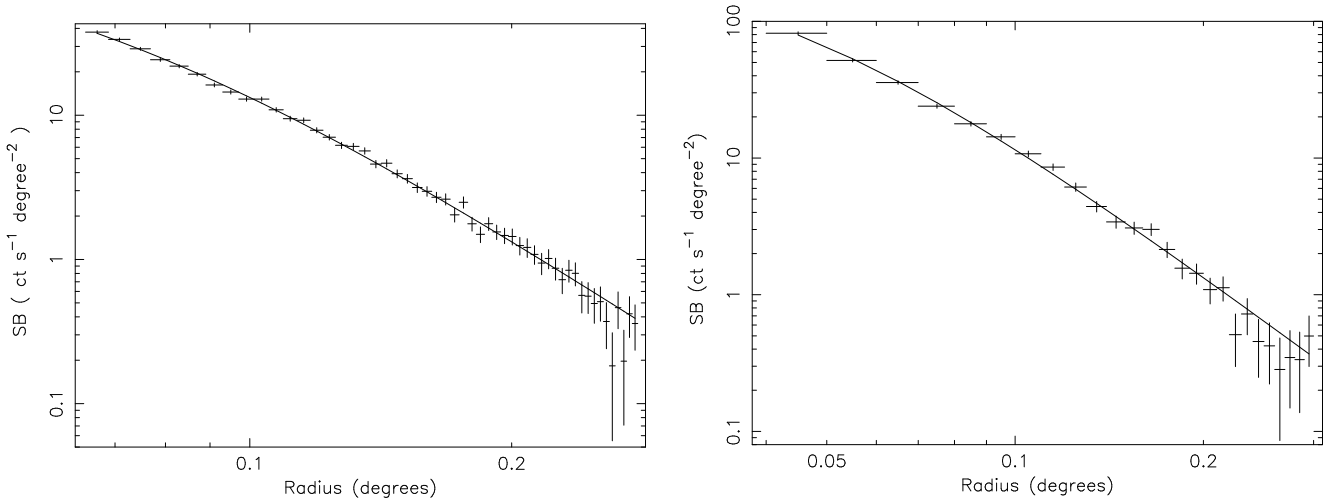

  \parbox[t]{1.0\textwidth}
  {\vspace{-0cm}\includegraphics[angle=270,width=8.5cm]{a1795_radial_qdp.eps}}
  \parbox[t]{1.0\textwidth}
  {\vspace{-6.6cm}\hspace{9cm}\includegraphics[angle=270,width=8.5cm]
    {a2029_radial_qdp.eps}} \vspace{-0cm} \hspace{8.5cm}
\caption{\label{fig:radial}
  Azimuthally averaged surface brightness profiles for A1795 (left panel)
  and A2029 (right panel). The solid line indicates the best-fitting model
  (see text for details).}
\end{figure*}

We obtained azimuthally-averaged surface brightness profiles for both
clusters, from \ROSAT\ PSPC data, in the following way. An image of the
cluster was extracted in the 0.2--2.4\keV\ band, and point sources above
4.5$\sigma$ significance were masked out. Using the master veto rate, the
contribution to the background from particles was subtracted, and the image
was then `flattened' by dividing by the corresponding exposure map, to
correct for the effects of vignetting. An estimate of the astrophysical
background was obtained, based on an annulus extracted from beyond the
cluster emission, with the PSPC support spokes removed.  Point sources were
also masked out from the annulus, and the remaining counts were
extrapolated across the field and subtracted from the source image.
Finally, a radial profile was extracted -- centred on the peak in the X-ray
emission -- in a series of fixed-width annuli, with a minimum of 50 counts
per bin (see Fig.~\ref{fig:radial}).

A King profile function (equation~\ref{eqn:2Dbeta}) was fitted to the data,
using the \textsc{qdp} package \citep{qdp_ug}, and all three parameters
were left free to vary. A small central region of the data was excised, to
prevent emission from the cooling flow biasing the results. The fitting was
repeated for a subset of the data, excluding emission beyond a fixed
radius, \rmsub{R}{outer}, so as to investigate any systematic variation in
$\beta$ with radius. The results of these fits are summarised in
table~\ref{tab:radfit}, and the best-fitting model to the whole image is
shown, together with the data points, in Fig.~\ref{fig:radial}.

It can be seen from Table~\ref{tab:radfit} that there is only marginal
evidence for a systematic trend in $\beta$ with radius for A2029, with a
large overlap between the confidence regions. However, A1795 seems to show
a significantly lower $\beta$ (0.70) for the innermost 9\arcmin, as
compared to the profile fitted out to 17\arcmin. An explanation of this
apparent flattening can be found in the core radius, which is significantly
smaller -- 2.2\arcmin, as compared to 3.7\arcmin\ for the whole profile.
This behaviour is an artefact of the excised central cooling region, which
is comparable in size to the gas core in the IGM. Consequently, \rc\ is
biased when a large part of the outer profile is excluded from the fit --
and this propagates through to the best-fitting $\beta$ value. This is
confirmed by the bottom row in Table~\ref{tab:radfit}, which lists the
results of re-fitting the innermost 9\arcmin\ with \rc\ frozen at the
best-fitting value from the full, 17\arcmin\ profile: the corresponding
$\beta$ (0.83) is, in fact, identical to the slope for the whole profile.
This problem is particularly pronounced for A1795 due to its fairly large
cooling flow: the excised central region amounts to $\sim$15 per cent of
\R200. The radius of the cooling flow in A2029 is only about 10 per cent of
\R200, a value which is more typical of groups, and hence the potential for
bias in \rc\ is minimal.

\citet{lew00} find a similar tendency for \rc\ and $\beta$ to decrease,
when fitting truncated radial profiles of simulated clusters. Similarly,
their analysis excludes a central portion of the data, corresponding to
emission from either the central cluster galaxy or a cooling flow. They
also claim to find a steepening in the logarithmic slope of the gas density
with radius in the outer regions, as reported by \citet{vik99}; however,
this discrepancy is only evident in the vicinity of the virial radius -- a
region rarely probed by observations of even the hottest clusters. In any
case, it is clear that, even if such a steepening is a significant effect,
it is not able to introduce large systematic \emph{relative} biases between
groups and clusters in our analysis.

\input{radfit_table} 

\section{Discussion}
\label{sec:discuss}
It is clear from the scaling properties we have examined, that virialized
systems do not exhibit self-similar behaviour. The $\beta-T$ relation and
the gas fraction data reveal a flattening of the gas density profiles,
which is most obvious in the group regime. These observations are
consistent with energy injection into the IGM by non-gravitational means.
However, three questions remain unanswered. Firstly, what caused this
heating; secondly, when did it take place and, thirdly, is self-similarity
broken only below a certain critical temperature, or does the transition
occur gradually?

There are three main candidates for the origin of the self-similarity
breaking. Both galaxy winds \citep[\eg][]{llo00} and AGN heating
\citep[\eg][]{wu00} are able to inject energy at roughly the levels
required to raise the entropy of gas in the central regions of the IGM.
However, the role of gas cooling in imposing an entropy floor
\citep[\eg][]{mua01} could also be significant. Although our results in
this paper provide no means of discriminating between the first two
options, our data do allow us to address the viability of cooling. The
cooling hypothesis is offered some support by our gas fraction results: the
variation in \fgas\ with \ewkT\ shown in figures~\ref{fig:fgasRV-kT} \&
\ref{fig:fgas_profiles}, is consistent with the loss of gas required in the
cool systems if cooling is to have a significant effect. On the other hand,
there is no evidence that groups have an excessive total \LB\ compared to
clusters, which would be expected if the cooling gas ultimately formed
stars \citep{hel02}. This may indicate that the \fgas\ trend is due to gas
being displaced to larger radius by the effects of energy injection, as
demonstrated by the simulations of \citet{met97}, for example. 

However, it should be noted that simulations which model the effects of
cluster mergers can generate a similarly extended gas distribution, without
energy injection from non-gravitational processes \citep{nav93,pea94}.
Similarly, while the large intrinsic scatter in the \srel{\beta}{T}
relation may reflect a difference in the level of energy injection between
haloes, it can also be attributed to the effects of hierarchical assembly:
\citet{cav99} predict a \srel{\beta}{T} relation with a 2$\sigma$ scatter
envelope, resulting from merging histories, which is in qualitative
agreement with the data in Fig.~\ref{fig:beta-kT}.

The second question raises two possibilities; either the heating took place
prior to halo collapse (so called `external' or `pre-'heating), or most
energy injection occurred after virialization (internal heating). Although
a rather simplistic distinction, these two scenarios will manifest
themselves in different ways on cluster properties, given the effect of
injecting energy into a medium in which a significant density gradient has
already been established. For example, preheating will tend to weaken the
shock boundary and move it outwards as compared to internal heating
\citep{toz00}. Since this boundary marks the point where the gas fraction
fades into the universal value, this amounts to an observable signature.
The evidence from \fgas\ is tentative, given the extrapolation
uncertainties, but there is indication of a systematic variation in the
total gas fraction within \R200 with \ewkT. Observations of the X-ray
background suggest that the heating phase took place over a time-scale of
$\sim$10$^{7}$~yr \citep{pen99}, although the epoch of energy injection is
not constrained.  Conversely, observations of the entropy floor
\citep{pon99,llo00} place a strict upper limit on \rmsub{z}{preh}, the
redshift at which the preheating epoch could have taken place, of
$\rmsub{z}{preh} \la 10$.

The question of on what mass scale the effects of self-similarity breaking
occur is more readily answered with our large sample. Whereas previous work
has pointed to a sharp transition between groups and clusters, our results
do not offer much support for this hypothesis. The analytical approach of
\citet{dos02} shows that the two mass regimes can be unified with a simple
model that incorporates energy injection from non-gravitational processes.
Their predicted scaling relations show a gradual steeping of \MT\ and \LT,
with decreasing temperature, and indicate that accretion shocks cannot be
completely suppressed in groups. Our measured \MT\ data offer some support
for a convex relation, as opposed to a broken power law, but the scatter is
rather large. Whilst higher quality data from \XMM\ and \Chandra\ will
doubtless shed some light on this issue, the greatest uncertainty lies in
the systematic bias associated with extrapolation to \R200: it is necessary
to trace the X-ray haloes of nearby groups out to \RV\ in order to resolve
the issue satisfactorily, and this calls for observations with a wider
field-of-view.

\subsection{Galaxies vs. groups}
\label{ssec:galVsgrp}
In a CDM Universe, the formation of structure proceeds in a bottom-up
fashion, with the smallest haloes virializing initially and subsequent
merging activity leading to the hierarchical assembly of progressively
larger haloes \citep[\eg][]{blu84}. Consequently, the smallest objects tend
to be older, having collapsed at an earlier epoch. This then leads to
differences in the scaling properties, as a result of the higher density of
the Universe at that time (\eg\ the \MT\ relation normalization). However,
in the context of a preheating prescription, invoked to explain the
breaking of self-similarity in galaxy groups, the timing of this early
formation epoch is critical; it is possible that these objects virialized
\emph{prior} to the preheating phase. This would give rise to behaviour
more consistent with massive clusters of galaxies, which are sufficiently
large as to be insensitive to the effects of energy injection. One possible
candidate for the origin of this preheating is population III stars, whose
formation precedes even that of galaxies, which may also have contributed
to the enrichment of the IGM \citep[\eg][]{loe01}.

As has already been seen, the properties of the galaxy-sized haloes in this
sample appear to differ from those of groups of a similar temperature.
Specifically, the gas density index ($\beta$) is rather steeper than
expected from a simple extrapolation of the $\beta-T$ relation for the
whole sample. In addition, the S0 galaxy NGC~1553 lies noticeably below the
\MT\ relation. This can be understood in terms of a large discrepancy
between its redshift of formation and observation: there is a bias towards
observing nearby objects of small mass, but these are likely to have formed
earliest of all sized haloes. For a given halo mass, the virial temperature
is proportional to $(1+\zf)$ and so, for NGC~1553, a value of $\zf
\sim3$--4 would increase \ewkT, and push it to the right of the \MT\ 
relation, as observed. Alternatively, the discrepancy can be attributed to
the effects of non-gravitational heating on \ewkT\ caused, for example, by
outflows from within the galaxy itself: a value of $\ewkT \simeq 0.3 \keV$
(compared to the actual value of 0.5\keV) would bring the point back on the
best-fitting \MT\ relation for the whole sample. This explanation is
further supported by its \LXLB, which is low enough for the stellar
population in the galaxy to have significantly influenced its X-ray halo
\citep[\eg][]{pel98}.

In fact, the behaviour of NGC~1553 is sufficiently unusual that it points
to an alternative formation mechanism to that which is usually invoked for
hot gas in groups and clusters. For example, the halo could have been built
from supernova-driven winds, originating in its own stellar population
\citep[\eg][]{cio91}, rather than from primordial material . This
explanation is supported by its \LXLB, and may explain the anomalously
steep $\beta$ index, which would otherwise be flattened by the effects of
energy injection prior to collapse.

In contrast, the position of the elliptical galaxy NGC~6482 on the \MT\ 
relation is fully consistent with the best fit line to the whole sample.
This is not surprising, as it is likely that this object is a fossil group
and hence will exhibit group-type behaviour. What is certain is that this
must be a very old system, given the long time-scale for orbital decay and
merging of the galaxy members. Although not as discrepant as NGC~1553,
NGC~6482 shows some suggestion that its $\beta$ index may be high for its
temperature. This could indicate that it formed sufficiently early to have
been relatively unaffected by a phase of energy injection, that would have
occurred prior to the formation of most larger objects. There is no doubt
that its halo must be predominantly primordial -- its \LXLB\ is sufficiently
high that the influence of its stellar population is negligible, in terms
of contributing to the total gas mass.

Given an early formation epoch, consistent with hierarchical formation,
coupled with a correspondingly larger mean density, it is particularly
important to consider the cooling time of the X-ray gas. At a radius of
0.1\R200, NGC~1553 has a gas cooling time (\tcool) of $\sim$6~Gyr and
NGC~6482 has $\tcool \simeq3$~Gyr. For comparison, the cooling times of the
next two coolest systems in our sample -- HCG~68 and NGC~1395 -- are 18~Gyr
and $\sim$25~Gyr, respectively. In the case of NGC~1553, this implies that
some form of heating mechanism must have prevented significant gas cooling,
if its halo was formed before $z = 1$ (corresponding to a light travel time
of roughly 6~Gyr). This is also supported by the fact that there appears to
be no evidence of strong cooling in the core, where the density is even
higher.  Energy injection from galaxy winds could provide the heating
mechanism necessary to explain this result, as suggested above. For
NGC~6482, some mechanism is also needed to prevent catastrophic gas
cooling, which is not observed even in the core, given that this fossil
group must be a very old system. Once again, its high \LXLB\ rules out a
significant heating contribution from supernova-driven winds. However,
there is some evidence of an AGN component in this galaxy \citep{gou94},
which may provide a suitable reheating mechanism to prevent the
establishment of a cooling flow.

\section{Conclusions}
We have studied the scaling properties of the X-ray emitting gas and
gravitating mass of a large sample of clusters, groups and galaxy-size
haloes. The 3-dimensional variations in gas density and
temperature -- corrected for contamination from cooling flows -- are
parametrized analytically, allowing us determine all derived quantities in
a self-consistent manner. We have derived virial radii and
emission-weighted temperatures from these models and are able to
extrapolate the properties of the data to measure them at fixed fractions
of \RV. We have also analysed an identical set of isothermal models, to
investigate the effects of neglecting spatial variations in the temperature
of the IGM.  We summarise our main findings below.

\begin{enumerate}
  \renewcommand{\theenumi}{(\arabic{enumi})}
\item Beta varies strongly with temperature, although there is evidence
  that galaxy-sized haloes do not follow this trend. We find a best fit
  power law relation of the form $\beta = (0.44\pm0.06) T^{0.20\pm0.03}$.
\item There is a $6\sigma$ correlation between \fgas\ within 0.3\R200\ and
  temperature, consistent with the variation in $\beta$. This trend is
  weakened only slightly (to $5.4\sigma$) by extrapolating the gas fraction
  to \R200 although, above 4\keV, the significance of this correlation
  drops to 3.2$\sigma$. The mean \fgas\ within \R200\ for the systems
  hotter than 4\keV\ is $(0.163 \pm 0.01)\h70^{-\frac{3}{2}}$, compared to
  $(0.134 \pm 0.01)\h70^{-\frac{3}{2}}$ for the whole sample. Under the
  assumption of isothermality, the scatter between \fgas\ at \R200\ and
  \ewkT\ is reduced, as is the normalization, giving a mean for the whole
  sample of $(0.110 \pm 0.01)\h70^{-\frac{3}{2}}$.
\item Observations of the variation in gas fraction as a function of radius
  in our sample reveal a systematic trend in gas fraction with temperature
  in all but the central regions (\la0.3\R200). This is consistent with the
  observed trend in \fgas\ with \ewkT.
\item In our study of the \MT\ relation, we employ two additional methods
  of calculating the average system temperature, one of which excludes the
  central region, another weighting the temperature with gas density rather
  than emissivity. We apply our three different methods within both
  0.3\R200\ and \R200, for both mass and \Tbar, to give a total of six \MT\ 
  relations. We find that the logarithmic slope of the relation is steeper
  within 0.3\R200 but that, even within \R200, it is inconsistent with
  self-similarity. There is close agreement between the measured slopes
  found for each of the three different prescriptions for \Tbar. For the
  emission-weighted \Tbar, within \R200, we find $M = 2.34\times 10^{13}
  \times T^{(1.84\pm 0.06)}$ \Msol. We find that the effect of assuming
  isothermality on the slope is negligible, but the normalization increases
  by 15 and 30 per cent for 0.3\R200\ and \R200, respectively
  \citep[\cf][]{neu99,hor99}, indicating that the total gravitating mass
  is significantly overestimated in our data when temperature gradients are
  neglected.  In addition, the scatter in the relation is reduced (and
  fully consistent with the parameter errors) compared to the
  non-isothermal case. The corresponding best fit relation is given by $M =
  3.02\times 10^{13} \times T^{(1.89\pm 0.04)}$ \Msol.
\item The relation between \R200\ and \ewkT, as deduced from simulated
  clusters \citep{nav95} deviates systematically from the measured values
  of \R200, as inferred from the overdensity profile. We find a strong
  negative correlation between the ratio of the NFW predicted \R200\ to
  our measured values and a quantitative measure of non-isothermality
  ($T(0)/T(0.3\R200)$). We show that only in the absence of a temperature
  gradient do the methods agree.
\item We address the issue of systematic bias associated with the
  extrapolation of the X-ray data to \R200, by fitting azimuthally averaged
  surface brightness profiles for two clusters, within different outer
  radii. We find no evidence for a significant variation in $\beta$ with
  cluster radius and conclude that the flatter gas density profiles of
  cooler systems cannot be attributed to the generally smaller angular
  range over which data are available for these objects.
\item We find that the two galaxies in the sample display unusual
  properties.  We have selected these objects on the basis of a lack of
  associated group or cluster halo emission, which can contaminate the
  galaxy halo flux.  The S0 galaxy, NGC~1553 has a steep $\beta$ index and
  falls to the right of the main \MT\ relation, indicative of an early
  formation epoch ($\zf \sim$ 3--4) -- which causes haloes of a given mass to
  be hotter than those collapsing at later times. It is also possible that
  \ewkT\ for this galaxy may have been artificially raised, probably by
  supernova-driven outflows from its stellar population. The elliptical
  galaxy NGC~6482 also shows a rather steep gas density profile, but
  otherwise exhibits group-like behaviour, consistent with a classification
  as a `fossil' group.
\end{enumerate}

\section*{Acknowledgments}
We thank Steve Helsdon for providing the software for measuring
non-statistical scatter, and Ewan O'Sullivan for useful discussions and
input. AJRS acknowledges financial support from the University of
Birmingham. This work made use of the Starlink facilities at Birmingham,
the LEDAS data base at Leicester and the NASA/IPAC Extragalactic Database
(NED).

\bibliography{$AJRS_LATEX/ajrs_bibtex} 
\label{lastpage}

\end{document}

%% file: target_table.tex

\begin{landscape}
\begin{longtable}[c]{l*{12}{c}}
\hline
Name & $z$ & T$^{a}$ & \R200\ & $\rho(0)$
& \rc\ & $\beta$ & $\alpha^{b}$ & $\gamma$ & CF radius$^{c}$ & Sample$^{d}$ & 
  Data$^{e}$\\[0.3mm]
  &  & (keV) & \mbox{~kpc} & ($\times 10^{-3}$cm$^{-3}$) & (arcmin) & &
 \keV/arcmin & & (arcmin) & & \\
\hline\hline\\[-3mm] 
\endhead
\hline
\multicolumn{13}{r}{\small \slshape continued overleaf} 
\endfoot
\hline\\[-6pt]
\endlastfoot
NGC 1553$\dagger$ & 0.0036 & $0.50^{+0.21}_{-0.13}$ & $203^{+90}_{-54}$ & $12.4^{+0.21}_{-2.25}$ & $1.04^{+0.30}_{-0.26}$ & $0.63^{+0.09}_{-0.07}$ &  -- & $ 1.44^{+0.15}_{-0.14}$ & -- & S & P \\ 
Virgo       & 0.0036 & $2.55^{+0.07}_{-0.06}$ & $1086^{+29}_{-30}$ & $60.26^{+0.24}_{-0.24}$ & $2.20^{+0.09}_{-0.09}$ & $0.45^{+0.02}_{-0.02}$ & $ -0.01^{+0.00}_{-0.00}$ & -- & 8.00 & F & P,S \\ 
NGC 1395    & 0.0057 & $0.84^{+0.24}_{-0.18}$ & $556^{+120}_{-95}$ & $14.1^{+12.6}_{-5.42}$ & $0.35^{+0.21}_{-0.19}$ & $0.43^{+0.03}_{-0.02}$ & -- & $ 1.05^{+0.07}_{-0.06}$ & 0.24 & S & P \\ 
NGC 5846   & 0.0058 & $1.18^{+0.07}_{-0.07}$ & $683^{+37}_{-33}$ & $57.18^{+0.67}_{-0.67}$ & $0.42^{+0.02}_{-0.02}$ & $0.55^{+0.02}_{-0.02}$ & -- & $ 1.06^{+0.01}_{-0.01}$ & 3.00 & F & P,S \\ 
HCG 68      & 0.0080 & $0.67^{+0.19}_{-0.15}$ & $497^{+104}_{-83}$ & $14.6^{+6.31}_{-4.22}$ & $0.37^{+0.12}_{-0.11}$ & $0.46^{+0.02}_{-0.02}$ & -- & $ 1.07^{+0.07}_{-0.06}$ & -- & L & P \\ 
NGC 5044    & 0.0090 & $1.25^{+0.06}_{-0.06}$ & $798^{+36}_{-33}$ & $10.66^{+3.45}_{-2.32}$ & $1.66^{+0.51}_{-0.28}$ & $0.49^{+0.01}_{-0.01}$ & -- & $ 0.97^{+0.02}_{-0.02}$ & 3.97 & L & P \\ 
NGC 3258    & 0.0095 & $2.57^{+0.12}_{-0.12}$ & $750^{+25}_{-25}$ & $2.32^{+0.02}_{-0.02}$ & $10.80^{+0.43}_{-0.43}$ & $0.32^{+0.01}_{-0.01}$ & -- & $ 1.60^{+0.07}_{-0.07}$ & 3.33 & F & P,S \\ 
IC 4296     & 0.0123 & $1.04^{+0.18}_{-0.15}$ & $529^{+57}_{-49}$ & $0.96^{+0.06}_{-0.06}$ & $2.64^{+0.11}_{-0.11}$ & $0.31^{+0.01}_{-0.01}$ & -- & $ 1.18^{+0.14}_{-0.14}$ & 2.27 & F & P,S \\ 
Abell 1060  & 0.0124 & $3.31^{+0.11}_{-0.1}$ & $1587^{+69}_{-40}$ & $3.74^{+0.02}_{-0.03}$ & $7.49^{+0.06}_{-0.01}$ & $0.72^{+0.01}_{-0.00}$ & -- & $ 0.97^{+0.01}_{-0.02}$ & 5.51 & L & P+G \\ 
NGC 6482$\dagger$ & 0.0131 & $0.56^{+0.37}_{-0.22}$ & $361^{+181}_{-105}$ & $25.0^{+13.5}_{-5.41}$ & $0.22^{\star}$ & $0.48^{+0.03}_{-0.04}$ & -- & $ 1.23^{+0.15}_{-0.15}$ & 1.20 & S & P \\ 
HCG 62      & 0.0137 & $1.48^{+0.18}_{-0.16}$ & $559^{+35}_{-31}$ & $1.25^{+0.02}_{-0.02}$ & $2.26^{+0.09}_{-0.09}$ & $0.30^{+0.01}_{-0.01}$ & -- & $ 1.46^{+0.08}_{-0.08}$ & 2.17 & F & P,S \\ 
Abell 262   & 0.0163 & $2.03^{+0.36}_{-0.27}$ & $998^{+146}_{-113}$ & $8.41^{+0.80}_{-0.73}$ & $1.49^{+0.17}_{-0.16}$ & $0.40^{+0.01}_{-0.01}$ & -- & $ 0.80^{+0.07}_{-0.08}$ & 0.41 & L & P \\ 
NGC 2563    & 0.0163 & $1.61^{+0.02}_{-0.03}$ & $627^{+6}_{-8}$ & $1.41^{+0.02}_{-0.01}$ & $2.07^{\star}$ & $0.42^{+0.003}_{-0.003}$ & -- & $ 1.36^{+0.01}_{-0.01}$ & 1.02 & S & P \\ 
NGC 507     & 0.0164 & $1.40^{+0.11}_{-0.08}$ & $738^{+28}_{-23}$ & $142.39^{+12.4}_{-9.91}$ & $0.10^{\star}$ & $0.43^{+0.01}_{-0.01}$ & $ 0.02^{+0.01}_{-0.01}$ & -- & 0.90 & L & P \\ 
IV Zw 038$^{1}$ & 0.0170 & $2.07^{+0.56}_{-0.42}$ & $892^{+104}_{-86}$ & $1.36^{+0.13}_{-0.11}$ & $2.77^{+0.40}_{-0.38}$ & $0.38^{+0.03}_{-0.03}$ & $ 0.04^{+0.03}_{-0.02}$ & -- & -- & L & P \\ 
AWM 7       & 0.0172 & $4.02^{+0.75}_{-0.62}$ & $2207^{+641}_{-420}$ & $5.22^{+0.05}_{-0.06}$ & $5.28^{+0.23}_{-0.07}$ & $0.59^{+0.00}_{-0.00}$ & -- & $ 0.67^{+0.09}_{-0.09}$ & 4.77 & L & P \\ 
Abell 194   & 0.0180 & $2.07^{+0.43}_{-0.43}$ & $1126^{+246}_{-199}$ & $0.66^{+0.02}_{-0.02}$ & $8.64^{+0.35}_{-0.35}$ & $0.60^{+0.02}_{-0.02}$ & $ -0.01^{+0.04}_{-0.04}$ & -- & 1.67 & F & P,S \\ 
MKW 4      & 0.0200 & $2.08^{+0.05}_{-0.06}$ & $842^{+18}_{-17}$ & $1.50^{+0.04}_{-0.04}$ & $5.45^{+0.22}_{-0.22}$ & $0.64^{+0.03}_{-0.03}$ & -- & $ 1.29^{+0.02}_{-0.02}$ & 1.51 & F & P,S \\ 
HCG 97      & 0.0218 & $1.00^{+0.13}_{-0.12}$ & $620^{+45}_{-37}$ & $140^{+205}_{-61}$ & $0.04^{+0.03}_{-0.01}$ & $0.41^{+0.01}_{-0.01}$ & $ 0.02^{+0.03}_{-0.03}$ & -- & -- & L & P \\ 
Abell 779   & 0.0229 & $3.57^{+0.94}_{-0.76}$ & $1075^{+203}_{-148}$ & $1.48^{+0.05}_{-0.05}$ & $1.42^{+0.06}_{-0.06}$ & $0.34^{+0.01}_{-0.01}$ & -- & $ 1.02^{+0.13}_{-0.14}$ & 5.26 & F & P,S \\ 
NGC 5129    & 0.0233 & $1.54^{+0.41}_{-0.35}$ & $567^{+71}_{-54}$ & $1.56^{+0.04}_{-0.04}$ & $2.46^{+0.10}_{-0.10}$ & $0.60^{+0.02}_{-0.02}$ & -- & $ 1.48^{+0.09}_{-0.10}$ & 3.85 & F & P,S \\ 
NGC 4325    & 0.0252 & $0.90^{+0.07}_{-0.07}$ & $678^{+81}_{-68}$ & $44.7^{+11.0}_{-15.0}$ & $0.21^{+0.07}_{-0.07}$ & $0.54^{+0.02}_{-0.01}$ & $ 0.00^{+0.02}_{-0.02}$ & -- & 0.78 & S & P \\ 
HCG 51      & 0.0258 & $1.38^{+0.04}_{-0.04}$ & $610^{+15}_{-15}$ & $0.96^{+0.02}_{-0.02}$ & $1.81^{+0.07}_{-0.07}$ & $0.30^{+0.01}_{-0.01}$ & $ -0.01^{+0.01}_{-0.01}$ & -- & 1.16 & F & H,S \\ 
NGC 6329    & 0.0276 & $1.60^{+0.52}_{-0.43}$ & $859^{+232}_{-153}$ & $1.17^{+0.07}_{-0.07}$ & $2.61^{+0.10}_{-0.10}$ & $0.53^{+0.02}_{-0.02}$ & -- & $ 1.06^{+0.16}_{-0.17}$ & 2.17 & F & P,S \\
NGC 6338    & 0.0282 & $2.64^{+1.92}_{-1.55}$ & $893^{+121}_{-278}$ & $4.44^{+0.68}_{-0.50}$ & $1.93^{+0.30}_{-0.26}$ & $0.53^{+0.04}_{-0.03}$ & -- & $ 1.25^{+0.09}_{-0.25}$ & 1.02 & S & P \\ 
MKW 4S      & 0.0283 & $2.46^{+0.23}_{-0.21}$ & $978^{+74}_{-65}$ & $1.42^{+0.03}_{-0.03}$ & $2.64^{+0.11}_{-0.11}$ & $0.51^{+0.02}_{-0.02}$ & -- & $ 1.14^{+0.06}_{-0.06}$ & 2.12 & F & P,S \\ 
Abell 539 & 0.0288 & $2.87^{+0.22}_{-0.21}$ & $1305^{+124}_{-104}$ & $2.42^{+0.06}_{-0.06}$ & $5.21^{+0.21}_{-0.21}$ & $0.69^{+0.03}_{-0.03}$ & -- & $ 1.04^{+0.06}_{-0.06}$ & -- & F & P,S \\ 
Klemola 44$^{2}$ & 0.0290 & $3.40^{+0.28}_{-0.26}$ & $1513^{+180}_{-148}$ & $4.74^{+0.05}_{-0.05}$ & $3.40^{+0.14}_{-0.14}$ & $0.61^{+0.02}_{-0.02}$ & -- & $ 0.94^{+0.06}_{-0.05}$ & -- & F & P,S \\ 
Abell 2199  & 0.0299 & $3.93^{+0.06}_{-0.06}$ & $1223^{+18}_{-15}$ & $12.09^{+0.01}_{-0.01}$ & $2.14^{0.001}_{0.001}$ & $0.60^{+0.0005}_{-0.0005}$ & -- & $ 1.15^{+0.01}_{-0.01}$ & 2.20 & L & P \\ 
Abell 2634  & 0.0309 & $3.45^{+0.28}_{-0.27}$ & $1189^{+104}_{-85}$ & $0.99^{+0.02}_{-0.02}$ & $8.62^{+0.34}_{-0.34}$ & $0.69^{+0.03}_{-0.03}$ & -- & $ 1.29^{+0.09}_{-0.09}$ & -- & F & P,S \\ 
AWM 4       & 0.0318 & $2.96^{+0.39}_{-0.39}$ & $1540^{+343}_{-290}$ & $3.52^{+0.08}_{-0.08}$ & $1.93^{+0.08}_{-0.08}$ & $0.62^{+0.02}_{-0.02}$ & $ -0.07^{+0.07}_{-0.07}$ & -- & 3.51 & F & P,S \\ 
Abell 496   & 0.0331 & $6.11^{+0.35}_{-0.43}$ & $1540^{+94}_{-69}$ & $7.24^{+0.31}_{-0.34}$ & $2.85^{+0.14}_{-0.12}$ & $0.62^{+0.01}_{-0.01}$ & -- & $ 1.16^{+0.03}_{-0.03}$ & 3.44 & L & P+G \\ 
2A0335+096  & 0.0349 & $3.34^{+0.3}_{-0.27}$ & $1596^{+158}_{-139}$ & $17.46^{+0.42}_{-0.42}$ & $1.40^{+0.06}_{-0.06}$ & $0.65^{+0.03}_{-0.03}$ & -- & $ 0.95^{+0.03}_{-0.03}$ & 2.63 & F & P,S \\ 
Abell 2052 & 0.0353 & $3.45^{+0.39}_{-0.4}$ & $1507^{+281}_{-237}$ & $10.03^{+0.17}_{-0.17}$ & $1.75^{+0.07}_{-0.07}$ & $0.64^{+0.03}_{-0.03}$ & $ -0.02^{+0.07}_{-0.07}$ & -- & 3.51 & F & P,S \\ 
Abell 2063 & 0.0355 & $4.00^{+0.12}_{-0.12}$ & $1493^{+57}_{-56}$ & $3.75^{+0.01}_{-0.01}$ & $3.79^{+0.15}_{-0.15}$ & $0.69^{+0.03}_{-0.03}$ & $ 0.05^{+0.02}_{-0.02}$ & -- & -- & F & P,S \\ 
Abell 3571  & 0.0397 & $7.31^{+0.28}_{-0.38}$ & $1870^{+101}_{-120}$ & $5.91^{+0.35}_{-.34}$ & $4.14^{+0.31}_{-0.31}$ & $0.69^{+0.01}_{-0.01}$ & -- & $ 1.12^{+0.04}_{-0.03}$ & 2.15 & M & P,G,S \\ 
MKW 9      & 0.0397 & $2.88^{+0.68}_{-0.55}$ & $1246^{+284}_{-212}$ & $4.86^{+0.11}_{-0.11}$ & $0.83^{+0.03}_{-0.03}$ & $0.52^{+0.02}_{-0.02}$ & -- & $ 0.97^{+0.09}_{-0.09}$ & 1.54 & F & I,S \\ 
Abell 2657 & 0.0400 & $4.53^{+0.61}_{-0.45}$ & $1251^{+188}_{-108}$ & $1.97^{+0.08}_{-0.08}$ & $5.68^{+0.31}_{-0.31}$ & $0.76^{+0.02}_{-0.02}$ & -- & $ 1.34^{+0.09}_{-0.12}$ & 2.15 & M & P,G,S \\ 
HCG 94      & 0.0417 & $4.02^{+0.46}_{-0.43}$ & $1151^{+94}_{-83}$ & $5.32^{+0.10}_{-0.10}$ & $1.10^{+0.04}_{-0.04}$ & $0.48^{+0.02}_{-0.02}$ & -- & $ 1.17^{+0.05}_{-0.05}$ & -- & F & P,S \\ 
Abell 119   & 0.0444 & $6.08^{+0.49}_{-0.47}$ & $1720^{+185}_{-135}$ & $1.68^{+0.02}_{-0.02}$ & $6.74^{+0.39}_{-0.39}$ & $0.66^{+0.02}_{-0.02}$ & -- & $ 1.14^{+0.08}_{-0.09}$ & 1.97 & M & P,G,S \\ 
MKW 3S     & 0.0453 & $4.42^{+0.57}_{-0.67}$ & $1218^{+176}_{-123}$ & $2.51^{+0.21}_{-0.20}$ & $4.13^{+1.38}_{-1.38}$ & $0.71^{+0.07}_{-0.07}$ & -- & $ 1.32^{+0.10}_{-0.11}$ & 2.05 & M & P,G,S \\ 
Abell 3558  & 0.0477 & $6.28^{+0.37}_{-0.3}$ & $1598^{+124}_{-87}$ & $5.94^{+0.08}_{-0.07}$ & $2.46^{+0.36}_{-0.36}$ & $0.55^{+0.03}_{-0.03}$ & -- & $ 1.13^{+0.04}_{-0.05}$ & 1.82 & M & P,G,S \\ 
Abell 4059  & 0.0480 & $5.50^{+0.5}_{-0.46}$ & $1313^{+161}_{-116}$ & $4.23^{+0.35}_{-0.33}$ & $2.85^{+0.65}_{-0.65}$ & $0.67^{+0.02}_{-0.02}$ & -- & $ 1.29^{+0.07}_{-0.08}$ & 1.82 & M & P,G,S \\ 
Tri. Aus.    & 0.0510 & $11.06^{+1.04}_{-0.96}$ & $1963^{+266}_{-188}$ & $4.85^{+0.15}_{-0.15}$ & $4.41^{+0.25}_{-0.24}$ & $0.67^{+0.01}_{-0.01}$ & -- & $ 1.26^{+0.08}_{-0.09}$ & 1.71 & M & P,G,S \\ 
Abell 85    & 0.0521 & $8.64^{+0.64}_{-0.29}$ & $1684^{+160}_{-61}$ & $3.56^{+0.11}_{-0.11}$ & $4.82^{+0.24}_{-0.24}$ & $0.76^{+0.02}_{-0.02}$ & -- & $ 1.32^{+0.03}_{-0.07}$ & 1.69 & M & P,G,S \\ 
Abell 3391  & 0.0536 & $5.39^{+0.72}_{-0.57}$ & $1671^{+306}_{-211}$ & $3.05^{+0.19}_{-0.18}$ & $2.44^{+0.12}_{-0.12}$ & $0.53^{+0.01}_{-0.01}$ & -- & $ 0.99^{+0.10}_{-0.11}$ & 1.63 & M & P,G,S \\ 
Abell 3266  & 0.0545 & $9.53^{+0.97}_{-0.55}$ & $1880^{+165}_{-103}$ & $2.85^{+0.03}_{-0.03}$ & $5.72^{+0.46}_{-0.46}$ & $0.74^{+0.04}_{-0.04}$ & -- & $ 1.29^{+0.05}_{-0.07}$ & 1.60 & M & P,G,S \\  
Abell 2319  & 0.0555 & $10.99^{+0.81}_{-1.14}$ & $1882^{+140}_{-113}$ & $7.45^{+0.11}_{-0.16}$ & $2.37^{+0.79}_{-0.79}$ & $0.54^{+0.06}_{-0.06}$ & -- & $ 1.23^{+0.04}_{-0.05}$ & 1.58 & M & P,G,S \\ 
Abell 780   & 0.0565 & $4.63^{+0.25}_{-0.24}$ & $2032^{+152}_{-133}$ & $10.09^{+0.25}_{-0.25}$ & $1.68^{+0.04}_{-0.04}$ & $0.67^{+0.01}_{-0.01}$ & -- & $ 0.90^{+0.03}_{-0.03}$ & 0.45 & L & P+G \\ 
Abell 2256  & 0.0581 & $8.62^{+0.55}_{-0.51}$ & $1814^{+124}_{-145}$ & $3.18^{+0.03}_{-0.03}$ & $5.02^{+0.11}_{-0.11}$ & $0.78^{+0.01}_{-0.01}$ & -- & $ 1.27^{+0.07}_{-0.05}$ & 1.53 & M & P,G,S \\ 
Abell 1795  & 0.0622 & $8.54^{+1.66}_{-1.05}$ & $2000^{+628}_{-290}$ & $4.30^{+0.05}_{-0.05}$ & $4.01^{+0.20}_{-0.21}$ & $0.83^{+0.02}_{-0.02}$ & -- & $ 1.17^{+0.10}_{-0.14}$ & 1.44 & M & P,G,S \\ 
Abell 3112  & 0.0703 & $7.76^{+1.65}_{-3.08}$ & $1311^{+237}_{-295}$ & $14.82^{+0.87}_{-0.86}$ & $1.03^{+0.69}_{-0.69}$ & $0.63^{+0.02}_{-0.02}$ & -- & $ 1.32^{+0.14}_{-0.09}$ & 1.29 & M & P,G,S \\ 
Abell 644   & 0.0711 & $11.68^{+1.52}_{-1.29}$ & $1660^{+299}_{-242}$ & $7.76^{+0.45}_{-0.43}$ & $2.18^{+0.18}_{-0.18}$ & $0.73^{+0.02}_{-0.02}$ & -- & $ 1.35^{+0.11}_{-0.10}$ & 1.27 & M & P,G,S \\ 
Abell 399   & 0.0722 & $7.97^{+0.69}_{-0.73}$ & $1734^{+149}_{-190}$ & $4.14^{+0.41}_{-0.41}$ & $1.89^{+0.36}_{-0.36}$ & $0.53^{+0.05}_{-0.05}$ & -- & $ 1.16^{+0.09}_{-0.06}$ & 1.26 & M & P,G,S \\ 
Abell 401   & 0.0739 & $9.55^{+0.45}_{-0.5}$ & $1851^{+113}_{-123}$ & $6.11^{+0.20}_{-0.20}$ & $2.37^{+0.09}_{-0.09}$ & $0.63^{+0.01}_{-0.01}$ & -- & $ 1.22^{+0.05}_{-0.04}$ & 1.23 & M & P,G,S \\ 
Abell 2670   & 0.0759 & $5.64^{+0.4}_{-0.39}$ & $1647^{+122}_{-111}$ & $6.20^{+0.16}_{-0.16}$ & $0.97^{+0.04}_{-0.04}$ & $0.55^{+0.02}_{-0.02}$ & -- & $ 1.04^{+0.04}_{-0.04}$ & -- & F & P,S \\ 
Abell 2029 & 0.0766 & $9.80^{+0.4}_{-0.42}$ & $2266^{+111}_{-103}$ & $6.34^{+0.10}_{-0.10}$ & $2.37^{+0.09}_{-0.09}$ & $0.68^{+0.03}_{-0.03}$ & $ 0.20^{+0.06}_{-0.06}$ & -- & 1.69 & F & P,S \\ 
Abell 1650  & 0.0845 & $8.04^{+1.75}_{-1.14}$ & $1816^{+756}_{-376}$ & $5.51^{+0.55}_{-0.55}$ & $2.25^{+0.78}_{-0.78}$ & $0.78^{+0.12}_{-0.12}$ & -- & $ 1.19^{+0.15}_{-0.17}$ & 1.09 & M & I,G,S \\ 
Abell 1651  & 0.0846 & $6.18^{+0.55}_{-0.36}$ & $1777^{+170}_{-116}$ & $6.25^{+0.41}_{-0.40}$ & $2.02^{+0.23}_{-0.23}$ & $0.70^{+0.02}_{-0.02}$ & -- & $ 1.10^{+0.04}_{-0.05}$ & 1.09 & M & P,G,S \\ 
Abell 2597  & 0.0852 & $6.02^{+0.47}_{-0.45}$ & $1841^{+161}_{-144}$ & $6.47^{+0.29}_{-0.29}$ & $1.40^{+0.06}_{-0.06}$ & $0.68^{+0.03}_{-0.03}$ & -- & $ 1.05^{+0.04}_{-0.04}$ & 1.55 & F & P,S \\  
Abell 478   & 0.0882 & $10.95^{+2.15}_{-1.82}$ & $1723^{+587}_{-332}$ & $6.98^{+0.21}_{-0.21}$ & $2.34^{+0.23}_{-0.23}$ & $0.75^{+0.01}_{-0.01}$ & -- & $ 1.34^{+0.17}_{-0.18}$ & 1.06 & M & P,G,S \\ 
Abell 2142  & 0.0894 & $11.16^{+1.54}_{-1.15}$ & $2216^{+544}_{-292}$ & $5.21^{+0.13}_{-0.13}$ & $3.14^{+0.22}_{-0.22}$ & $0.74^{+0.01}_{-0.01}$ & -- & $ 1.18^{+0.10}_{-0.13}$ & 1.05 & M & P,G,S \\ 
Abell 2218  & 0.1710 & $8.28^{+1.82}_{-1.33}$ & $1904^{+180}_{-149}$ & $6.17^{+0.15}_{-0.16}$ & $0.90^{\star}$ & $0.59^{+0.01}_{-0.01}$ & -- & $ 1.11^{+0.02}_{-0.02}$ & -- & L & P+G \\ 
Abell 665   & 0.1818 & $8.60^{+1.27}_{-0.94}$ & $2273^{+268}_{-279}$ & $6.32^{+0.17}_{-0.16}$ & $1.21^{+0.04}_{-0.04}$ & $0.65^{+0.01}_{-0.01}$ & -- & $ 1.02^{+0.08}_{-0.06}$ & -- & L & P+G \\ 
Abell 1689  & 0.1840 & $12.31^{+1.19}_{-0.93}$ & $2955^{+135}_{-112}$ & $33.61^{+0.64}_{-1.92}$ & $0.60^{+0.02}_{-0.00}$ & $0.73^{+0.06}_{-0.00}$ & $ 0.00^{\star}$ & -- & 2.40 & L & P+G \\ 
Abell 2163  & 0.2080 & $16.64^{+3.36}_{-1.55}$ & $2104^{+794}_{-253}$ & $7.77^{+0.53}_{-0.52}$ & $1.63^{+0.08}_{-0.08}$ & $0.73^{+0.02}_{-0.02}$ & -- & $ 1.38^{+0.11}_{-0.22}$ & 0.54 & M & P,G,S \\ 
\hline
\\*\\* 
\caption{
  Some key properties of the 66 objects in the sample, listed in order of
  increasing redshift. Redshifts are taken from
  \citet{ebe96,ebe98,pon96} and NED. Columns 3--9 are data as determined
  in this work. All errors are 68\% confidence.\\*\\[-3mm]
  $\dagger$ indicates the two galaxies. \\
  $^{\star}$ denotes no errors available, as parameter poorly constrained.\\
  $^{1}$ also known as NGC~383. \\
  $^{2}$ also known as Abell~4038. \\[-2mm]
  \newline $^{a}$ The cooling-flow corrected, emission-weighted temperature
  of the system within 0.3\R200, as determined in this work.  \newline
  $^{b}$ Temperature gradient; positive values mean $T$ \emph{decreases}
  with radius.  \newline $^{c}$ Cooling flow excision radius (M sample) or 
  radius within which a cooling flow component was fitted (F,L,S samples) 
  \newline $^{d}$ F = Finoguenov \etal, L = Lloyd-Davies
  \etal, M = Markevitch \etal, S = Sanderson \etal\ (this work) \newline
  $^{e}$ P = \ROSAT\ PSPC, H = \ROSAT\ HRI, G = \ASCA\ GIS, S = \ASCA\ SIS,
  I = \Einstein\ IPC; + denotes simultaneous fit }
\label{tab:sample}
\end{longtable}
\end{landscape}

%% file: MT_fit_table.tex

\begin{table*}
\begin{tabular}{lcccccc}
\hline\\[-3mm] 
kT weighting & Integration radius & Index & Normalization &
Scatter$^{a}$ & Mean$^{b}$ & $\sigma^{c}$ \\[-0.5mm]
       & (\R200) &       & (log(\Msol))       & \\
\\[-3mm]
\hline\hline\\[-3mm] 
\textit{Non-isothermal} \\
A....Emission & 0.3 & $1.92\pm0.06$ & $12.80\pm0.03$ & $1.02$ & -- & -- \\
B....Emission (CF excised) & 0.3 & $1.94\pm0.06$ & $12.80\pm0.03$ & $0.98$ &
     0.98 & 0.04 \\
C....Mass & 0.3 & $1.97\pm0.06$ & $12.85\pm0.03$ & $0.78$ & 0.93 & 0.14 \\
\textbf{D....Emission} & \textbf{1.0} & $\mathbf{1.84\pm0.06}$ & $\mathbf{13.37\pm0.03}$ & 
\textbf{0.96} & \textbf{0.92} & \textbf{0.12} \\
E....Emission (CF excised) & 1.0 & $1.86\pm0.06$ & $13.37\pm0.03$ & $0.90$ &
     0.91 & 0.15 \\
F....Mass & 1.0 & $1.83\pm0.06$ & $13.58\pm0.03$ & $0.52$ & 0.78 & 0.41 \\
\hline\\[-3mm] 
\textit{Isothermal} \\
A$^{\prime}$...Emission & 0.3 & $1.97\pm0.05$ & $12.86\pm0.03$ & $0.89$ & -- & -- \\
B$^{\prime}$...Emission (CF excised) & 0.3 & $1.98\pm0.05$ & $12.86\pm0.03$ & $0.89$ & --- & --- \\
C$^{\prime}$...Mass & 0.3 & $2.02\pm0.06$ & $12.85\pm0.03$ & $0.83$ & --- & --- \\
D$^{\prime}$...Emission & 1.0 & $1.89\pm0.04$ & $13.48\pm0.02$ & $0.65$ & --- & --- \\
E$^{\prime}$...Emission (CF excised) & 1.0 & $1.90\pm0.04$ & $13.48\pm0.02$ & $0.64$ & --- & --- \\
F$^{\prime}$...Mass & 1.0 & $1.87\pm0.05$ & $13.52\pm0.02$ & $0.44$ & --- & --- \\
\hline                                      
\end{tabular}
\caption{
\label{tab:MTfit}
      Summary of results for the
      power law \MT\ fitting using different mean temperature 
      prescriptions and integration radii (for measuring both mass and 
      temperature). Primed models in the lower half of the table are
      \emph{isothermal} models, which have identical mean temperatures to 
      models A-F, but different total masses. The bold row indicates our 
      default \MT\ relation. Notes:
      $^{a}$Multiples of the statistical scatter expected from 
      the errors alone.
      $^{b}$denotes the mean ratio of kT divided by the corresponding
      values obtained with prescription A and $^{c}$ is the standard 
      deviation of these ratios across the sample (these numbers have been 
      omitted from the lower half since the ratios depend only on temperature, 
      which is unchanged).
         }
\end{table*}

%% file: radfit_table.tex

\begin{table*}
\begin{tabular}{lccccccc}
\hline\\[-3mm] 
Cluster name & Temperature & \rmsub{R}{outer} & \rmsub{R}{outer}/\R200\ & Normalization & $\beta$ & \rc\ & red.~\chisq/dof \\[-0.5mm]
 & (keV) & (arcmin) &      & (ct s$^{-1}$/degree$^{2}$) &  & (arcmin) & \\
\\[-3mm]
\hline\hline\\[-3mm] 
A2029 & 9.80 & 18 & 0.68 & $294^{+9}_{-33}$    & $0.73^{+0.02}_{-0.01}$ &
$2.51^{+0.20}_{-0.20}$    & 1.25/23 \\
      &      & 12 & 0.45 & $356^{+74}_{-53}$   & $0.70^{+0.02}_{-0.02}$ &
$2.17^{+0.24}_{-0.24}$ & 1.35/13 \\
      &      &  9 & 0.34 & $399^{+130}_{-77}$  & $0.68^{+0.03}_{-0.03}$ &
$1.99^{+0.31}_{-0.32}$ & 1.72/8  \\
A1795 & 8.54 & 17 & 0.60 & $173^{+21}_{-17}$   & $0.83^{+0.02}_{-0.02}$ &
$3.72^{+0.25}_{-0.24}$ & 1.04/50 \\
      &      & 12 & 0.42 & $209^{+41}_{-30}$   & $0.79^{+0.03}_{-0.03}$ &
$3.26^{+0.33}_{-0.33}$ & 1.18/30 \\
\hbox{}~~~~~~~~~~$\dagger$  &      & 9 & 0.32 & $401^{+364}_{-130}$ & $0.70^{+0.04}_{-0.03}$ &
$2.18^{+0.53}_{-0.59}$ & 1.09/18 \\
\rc\ fixed$^{\star}$ &      &  9 & 0.32 & $169^{+5}_{-5}$  & $0.83^{+0.01}_{-0.01}$ &
$3.72$ & 1.49/19 \\
\hline                                      
\end{tabular}
  \center{\caption{\label{tab:radfit}Summary of results for the
      1-dimensional surface brightness fitting within different
      radii. Errors are 1$\sigma$. A central region of the data 
      was excluded to avoid contamination of by cooling flow 
      emission; the radii of exclusion were 2.7\arcmin\ \& 
      4\arcmin\ for A2029 \& A1795, respectively. $\dagger$This fit 
      was noticeably 
      biased by the excised central region of the data and was repeated: 
      $^{\star}$ indicates that the core radius was frozen at its 
      previously-determined best-fit value to the whole profile 
      (see text for details).}}
\end{table*}